\documentclass[prb,reprint,superscriptaddress,aps,showpacs,showkeys,floatfix,twocolumn,longbibliography]{revtex4-1}
\usepackage{keyval} \usepackage{dcolumn}
\usepackage{times,xspace}
\usepackage{amsbsy,amssymb,bm,mathptmx}
\usepackage{graphicx,color,epsfig,rotate}
\usepackage{fancyhdr}
\usepackage{amsmath,amssymb,amsfonts,amsthm}
\usepackage{tabularx,siunitx}
\usepackage{booktabs}
\usepackage[]{units}
\usepackage{cancel}
\usepackage{xcolor}
\usepackage{graphicx}
\usepackage{dsfont}

\renewcommand{\vec}[1]{\bm{#1}}
\newcommand{\mat}[1]{\overline{\bm{#1}}}
\newcommand{\abs}[1]{\left| #1 \right|}
\newcommand{\myvec}[3]{\begin{pmatrix}#1\\#2\\#3\end{pmatrix}}
\newcommand{\vc}[1]{\mathbf{#1}}
\begin{document}
\title{Search for the magnetic monopole at a magnetoelectric surface}

\author{Q. N. Meier$^*$}
\affiliation{Department of Materials, ETH Zurich, CH-8093 Z\"urich, Switzerland}
\author{M. Fechner$^*$}
\affiliation{Department of Materials, ETH Zurich, CH-8093 Z\"urich, Switzerland}
\affiliation{Max Planck Institute for the Structure and Dynamics of Matter, 22761 Hamburg, Germany}
\author{T. Nozaki}
\affiliation{Department of Electronic Engineering, Tohoku University, Sendai 980-8579, Japan}
\author{M. Sahashi}
\affiliation{Department of Electronic Engineering, Tohoku University, Sendai 980-8579, Japan}
\affiliation{ImPACT Program, Japan Science and Technology Agency, Tokyo 102-0076, Japan}
\author{Z. Salman}
\affiliation{Laboratory for Muon Spin Spectroscopy, Paul Scherrer Institute, CH-5232 Villigen PSI, Switzerland}
\author{T. Prokscha}
\affiliation{Laboratory for Muon Spin Spectroscopy, Paul Scherrer Institute, CH-5232 Villigen PSI, Switzerland}
\author{A. Suter}
\affiliation{Laboratory for Muon Spin Spectroscopy, Paul Scherrer Institute, CH-5232 Villigen PSI, Switzerland}
\author{P. Schoenherr}
\affiliation{Department of Materials, ETH Zurich, CH-8093 Z\"urich, Switzerland}
\author{M. Lilienblum}
\affiliation{Department of Materials, ETH Zurich, CH-8093 Z\"urich, Switzerland}
\author{P. Borisov}
\affiliation{Department of Physics, School of Science, Loughborough University, Loughborough, LE11 3TU, UK}
\author{I. E. Dzyaloshinskii}
\affiliation{School of Physical Sciences, University of California Irvine, Irvine, CA 92697, USA}
\author{M. Fiebig}
\affiliation{Department of Materials, ETH Zurich, CH-8093 Z\"urich, Switzerland}
\author{H. Luetkens}
\affiliation{Laboratory for Muon Spin Spectroscopy, Paul Scherrer Institute, CH-5232 Villigen PSI, Switzerland}
\author{N. A. Spaldin}
\affiliation{Department of Materials, ETH Zurich, CH-8093 Z\"urich, Switzerland}

\date{\today}
\begin{abstract}
We show, by solving Maxwell's equations, that an electric charge on the surface of a slab of a linear magnetoelectric material generates an image magnetic monopole below the surface provided that the magnetoelectric has a diagonal component in its magnetoelectric response. The image monopole, in turn, generates an ideal monopolar magnetic field outside of the slab. Using realistic values of the electric- and  magnetic- field susceptibilities, we calculate the magnitude of the effect for the prototypical magnetoelectric material Cr$_2$O$_3$. We use low energy muon spin rotation to measure the strength of the magnetic field generated by charged muons as a function of their distance from the surface of a Cr$_2$O$_3$ film, and show that the results are consistent with the existence of the monopole. We discuss other possible routes to detecting the monopolar field, and show that, while the predicted monopolar field generated by Cr$_2$O$_3$ is above the detection limit for standard magnetic force microscopy, detection of the field using this technique is prevented by surface charging effects. \\

\noindent
$^*$ These authors contributed equally to this work
\end{abstract}

\maketitle

\section{introduction}
The elusiveness of magnetic monopoles, which are expected in classical electrodynamics because of the duality symmetry between electricity and magnetism, has intrigued physicists for centuries. Their relevance was particularly emphasized by Dirac, who introduced a description allowing monopoles to remain consistent with the known zero divergence of magnetic fields, and showed that their existence would explain the observed quantization of electric charge in the universe\cite{Dirac:1931vc}. The quest for a magnetic monopole therefore remains an active research area today, ranging from searches using sensitive cosmic-ray detectors to attempts to generate monopoles in collider experiments; for a review see Ref.~\onlinecite{Rajantie:2016}. 
While the existence of true magnetic monopoles has not yet been verified, a number of condensed-matter systems have been shown to provide intriguing analogues. Perhaps the most popular are the pyrochlore-structure ``spin-ice'' materials of which the prototype is dysprosium titanate, Dy$_2$Ti$_2$O$_7$\cite{Castelnovo:2008hb,Morris:2009kh}. In these materials, magnetic excitation of the frustrated antiferromagnetic ``two-in, two-out'' tetrahedral spin ordering leads to two locally divergent magnetizations of opposite sign -- one tetrahedron has three spins pointing inward and one pointing outward, and vice versa -- connected by the analogue of a Dirac string. Also of interest are the so-called linear magnetoelectric materials, magnetic insulators in which an applied electric field induces a magnetization and vice versa. Here, it has been shown theoretically that when an electric charge is introduced into a diagonal magnetoelectric (in which the induced magnetization is parallel to the electric field), the divergent electric field of the charge induces a monopole-like magnetization around the electric charge\cite{Khomskii:2014dp,Fechner:2014us}. Similarly, it has been argued that a charge above a topological insulator/ferromagnet heterostructure should lead to a magnetic monopolar field due to the quantized Chern-Simons magnetoelectric response of topological insulators with broken time-reversal symmetry\cite{Qi:2008eu,Qi:2009ip}. While the magnetoelectric response of such a system can in principle be sizable\cite{Coh_et_al:2011}, its detection is challenging\cite{Salman_et_al:2018} because of the practical difficulty in achieving insulating bulk behavior in topological insulators, as well as the need to incorporate a separate time-reversal symmetry breaking component\cite{Pesin/MacDonald:2013}.

\begin{figure}
\centering
\includegraphics[scale=0.34]{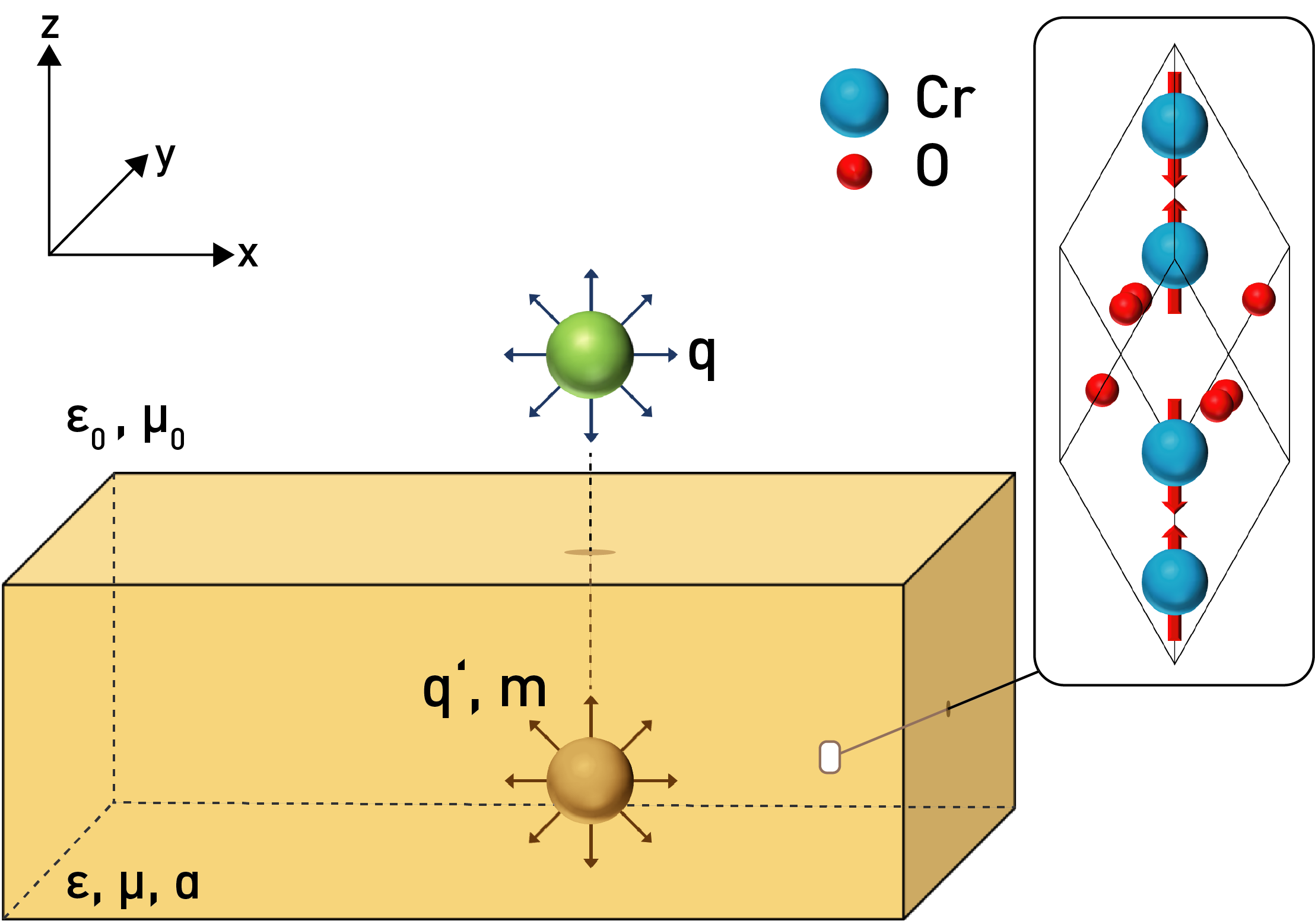}
\caption{\label{fig1} A charge $q$ (green sphere) above the surface of a magnetoelectric induces an image monopole $m$ (brown sphere) at the same distance beneath the surface. The magnetic field above the surface is divergent with its source at the subsurface image monopole. The zoom shows the unit cell of the prototypical magnetoelectric Cr$_2$O$_3$.}
\end{figure}

Here we show that conventional linear magnetoelectric materials, of which Cr$_2$O$_3$ is the prototype\cite{dzyalo, astrov}, can generate an external monopolar magnetic field when an electric charge is placed above any flat sample surface. 
In linear magnetoelectrics, an applied magnetic field $\vec{H}$ induces an electric polarization $\vec{P}$ and an applied electric field $\vec{E}$ induces a magnetization $\vec{M}$ according to
\begin{align}
\vec{P} & = \mat{\alpha} \vec{H} \\
\mu_0\vec{M} & = \mat{\alpha}^T \vec{E}  \quad.
\end{align}
Here $\mu_0$ is the permability of free space and $\mat{\alpha}$ is the magnetoelectric tensor in SI units. $\mat{\alpha}$ is allowed to be non-zero in materials that break both time-reversal and space-inversion symmetry, and its non-zero components are determined by the detailed crystalline and magnetic symmetry.
In the next section, we show theoretically that, in cases for which $\mat{\alpha}$ has a non-zero diagonal component, a surface charge $q$ generates a sub-surface image monopole $m$. This leads in turn to a divergent magnetic field above the sample surface.

\section{Calculation of the fields induced by a charge on the surface of a magnetoelectric material}
\label{Section2}

We consider the geometry shown in Fig. \ref{fig1}, in which a point electric charge $q$ is placed in the vacuum region a small distance $\vec r_0=(0,0,z_0)$ away from the planar surface of a semi-infinite slab of a uniaxial magnetoelectric material. 

\subsection{Magnetoelectrostatics}\label{chap:mestat}
We solve the classical Maxwell equations for a static system in which the electromagnetic fields are given by Gauss' laws 
\begin{align}
\nabla\cdot\vec{D}&=\rho\label{eqn::EG}\\
\nabla\cdot\vec{B}&=0\label{eqn::BG}\quad .
\end{align}
Here $\rho$ is the free charge, $\vec{D}$ is the electric displacement and $\vec{B}$ is the magnetic flux density. In the conventional treatment the latter are given by $\vec{D}=\mat{\epsilon} \vec{E}$ and $\vec{B}=\mat{\mu} \vec{H}$ respectively, with $\mat\epsilon$ and $\mat\mu$ the dielectric and magnetic susceptibility. Inside a linear magnetoelectric material, however, the displacement and magnetic fields take the form
\begin{eqnarray}
\vec{D}&=\mat{\epsilon} \vec{E} &+ \mat{\alpha} \vec{H} \\
\vec{B}&=\mat{\mu} \vec{H} &+ \mat{\alpha}^T \vec{E} \quad,
\end{eqnarray}
where $\mat\alpha$ is the linear magnetoelectric susceptibility tensor \cite{Fiebig:2005} and $\mat{\alpha}^T$ its transpose. This expanded formulation must be used in the Maxwell equations \eqref{eqn::EG} and \eqref{eqn::BG} to calculate the electromagnetic fields in a magnetoelectric material. 
In addition, the system needs to satisfy the electrostatic boundary conditions for interfaces at all times: 

\begin{align}
\vec{D}\cdot \vec{n} &= \text{constant}\\
\vec{B}\cdot\vec{n}&= \text{constant}\\
\vec{E}\cdot \vec{t} &= \text{constant}\\
\vec{H}\cdot\vec{t}&= \text{constant}\quad ,
\end{align}
where $\vec{n}$ is the surface normal and $\vec{t}$ the surface tangent.

Since we look at the static limit, it is helpful to use the electrostatic and magnetostatic potentials, $\phi_e$ and $\phi_m$, which are related to the electric and magnetic fields by 
\begin{align}
\vec{E}&=-\nabla\phi_e\\
\vec{H}&=-\nabla\phi_m\quad .
\end{align}

\subsection{Solution for an isotropic magnetoelectric}

\begin{figure}
\includegraphics[scale=0.95]{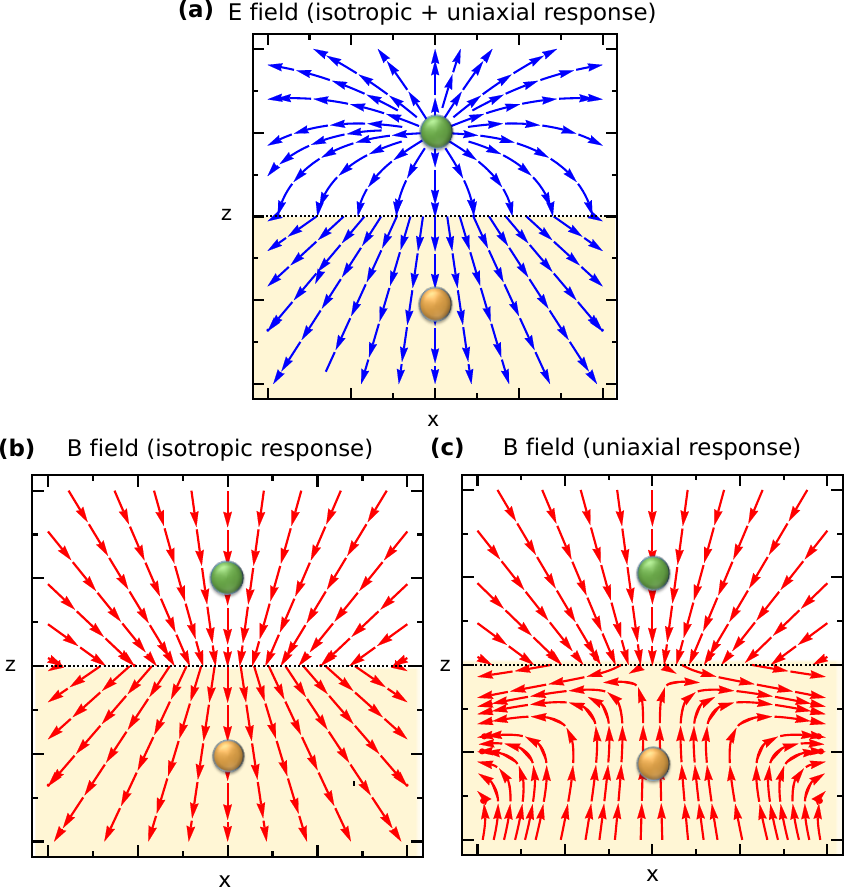}
\caption{Calculated $\vec{E}$ (a) and $\vec{B}$ (b)-(c) fields induced by a  positive charge close to a magnetoelectric surface in Cr$_2$O$_3$. The green and orange points mark the position of the charge and the image charges. The arrows indicate the orientation of the fields (b) is calculated using a isotropic magnetoelectric response and (c) using the full anisotropic responses of Cr$_2$O$_3$. The electric field (a) is indistinguishable for the two cases due to the small anisotropy in the dielectric response.   }\label{fig:plotfield}
\end{figure}

First, we present the solution of the field equations for a charge above an isotropic magnetoelectric in which $\mat{\alpha}=\alpha\mathds{1}$ ($\mathds{1}$ is the unit matrix). Even though there are five magnetic point groups permitting such behavior, no material with such a magnetoelectric response has yet been identified  experimentally. Nevertheless, the behavior is of academic interest, since it has the symmetry of the so-called Chern-Simons magnetoelectric response of topological insulators \cite{Coh/Vanderbilt:2014}. In addition, the solution is obtained straightforwardly using the well-established method of mirror charges, and already provides insight into the full problem that we address in the next section.  Placing mirror charges inside, $in$, and outside, $out$, of the magnetoelectric we obtain the 
ansatz for the electric potential, $\phi_e$:
\begin{eqnarray}
\phi_e^{out} (\vec{r})&=& \dfrac{1}{4\pi\epsilon_0}\frac{q}{\abs{\vec{r}-\vec{r_0}}}+\frac{q'}{\abs{\vec{r}-\vec{r_1}}} \\
\phi_e^{in} (\vec{r})&=& \frac{q''}{\abs{\vec r- \vec r_0}} \quad ,
\label{eq::pot1}
\end{eqnarray}

where $q$ is the real charge at position $\vec r_0=(0,0,z_0)$, and $q''$ and $q'$ are electric image charges at positions $\vec r_0=(0,0,z_0)$, $\vec r_1=(0,0,-z_0)$.
We enforce continuous normal components of the displacement field and magnetic flux density at the interface as well as continuous tangential components of the electric and magnetic fields. To satisfy the magnetic boundary conditions, we use the following ansatz for the magnetic potential:
\begin{eqnarray}
\phi_m^{out} (\vec{r}) &=& \frac{m'}{\abs{\vec{r}-\vec{r_1}}} \\
\phi_m^{in} (\vec{r}) &=& \frac{m''}{\abs{\vec r- \vec r_0}} \quad ,
\label{eq::pot2}
\end{eqnarray}
where $m''$ and $m'$ are effective magnetic image monopoles at positions $\vec r_0=(0,0,z_0)$, $\vec r_1=(0,0,-z_0)$. We solve this system of equations, as shown in detail in Appendix~\ref{app:iso}, to obtain the following expression for the magnetic flux density outside of the material: 
\begin{equation}\label{eq::iso}
\vec B(\vec r)=-\dfrac{\mu_0}{4\pi}\dfrac{2q\alpha}{(\mu+\mu_0)(\epsilon+\epsilon_0)-\alpha^2}\dfrac{\vec r-\vec r_1}{|\vec r - \vec r_1|^{3}}\quad .
\end{equation}
The resulting $\vec{E}$ and $\vec{B}$ fields both inside and outside of the magnetoelectric slab, using literature values for the relative response parameters of Cr$_2$O$_3$ (Table~\ref{tab:parameters}) averaged to mimic an isotropic material, are sketched in  Figs.~\ref{fig:plotfield} a) and b).

The electric field outside the slab is similar to that of the original isolated point charge, with deviations in the region close to the interface due to the dielectric screening of the field within the slab.
The electric field within the slab is a divergent point charge field with the charge outside the slab as its origin, and its magnitude screened by the static dielectric constant of the material.
Since the magnetic flux density within the material is induced by the electric field through the magnetoelectric effect, the field lines within the slab diverge identically to those of the electric field. Outside of the slab, the magnetic field is particularly interesting as it is perfectly divergent, with its source being an image monopole that is the same distance below the surface as the point charge is above  it. A positive charge with the magnitude of an electronic charge induces an image monopole = \unit[-3.63$~10^{-16}$]{Am} in a material with these response parameters. This converts to a magnetic $\vec{B}$-field of the order of a $\mu$T caused by and measured at the site of a single electronic point charge placed a distance of 2 nm above the interface. Note that a positive charge on a material with a positive magnetoelectric tensor induces a \textit{negative} magnetic field outside the material, and that changing the sign of one of the surface charge or the magnetoelectric tensor changes the sign of the field. As a result, opposite magnetoelectric domains produce fields of opposite sign for the same sign of charge.

\begin{table}
\begin{ruledtabular}
\begin{tabular}{c|rcc}
component & $\alpha$ (ps/m)  & $\epsilon_r$ ($\epsilon_0$)   & $\mu_r$ ($\mu_0$) \\ \hline
$\perp$   & 0.734 & 10.3   & 1.0014  \rule[-1ex]{0pt}{3.5ex}  \\
$\|$         & -0.233 & 10.9  & 1.0001  \rule[-1ex]{0pt}{3.5ex}  \\
\end{tabular}
\caption{Experimental values of $\alpha$, relative permittivity $\epsilon_r$ and relative permeability $\mu_r$ for Cr$_2$O$_3$, from Refs.~\onlinecite{Foner:1963vi,Wiegelmann:1994tq,LAL:1967fd}. $\epsilon_r$ is measured at room temperature, whereas $\mu_r$ and $\alpha$ are the low temperature (4 K) values. $\alpha$ has units of inverse velocity in the SI units that we use here.}
\label{tab:parameters}
\end{ruledtabular}
\end{table}

\subsection{Solution for a uniaxial magnetoelectric}

Next we analyze the realistic case of the response of a uniaxial anisotropic magnetoelectric material\footnote{While only 11 of the 58 magnetic point groups that allow the magnetoelectric effect have uniaxial symmetry, in many of the other cases, the tensor can be transformed into a form with a diagonal component and our analysis remains relevant for an appropriate choice of surface cut.}. Specifically, we take the case of the prototypical magnetoelectric, Cr$_2$O$_3$, and treat its full uniaxial response. We orient the high-symmetry axis along the $z$ axis, so that the magnetoelectric, dielectric and magnetic susceptibility tensors are as follows: 

\scalebox{0.9}{
$\mat{\alpha}= \begin{bmatrix} \alpha_\perp &0&0\\ 0&\alpha_\perp&0 \\ 0&0&\alpha_\|
\end{bmatrix}, \quad  \mat{\epsilon}= \begin{bmatrix} \epsilon_\perp &0&0\\ 0&\epsilon_\perp&0 \\ 0&0&\epsilon_\|
\end{bmatrix}, \quad \mat{\mu}= \begin{bmatrix} \mu_\perp &0&0\\ 0&\mu_\perp&0 \\ 0&0&\mu_\|
\end{bmatrix} \quad .$}\\

Aligning the $\vec{n}=(0,0,z)$ axis of the magnetoelectric perpendicular to the surface plane, the field equations inside the magnetoelectric  become
\begin{align} 
\nabla\cdot \vec{D} &=(\epsilon_\perp\nabla_\perp + \epsilon_\parallel\nabla_\parallel)\vec{E}+(\alpha_\perp\nabla_\perp+
\alpha_\parallel\nabla_\parallel)\vec{H}=0\nonumber \\
\nabla \cdot \vec{B} &=(\mu_\perp \nabla_\perp+\mu_\parallel\nabla_\perp)\vec{H}+(\alpha_\perp \nabla_\perp+\alpha_\parallel\nabla_\parallel)\vec{E}=0\quad ,
\end{align}
and those outside the material 
\begin{align}
\nabla\cdot\vec D &=q\delta(\vec r - \vec r_0) \nonumber \\
\nabla\cdot\vec B &= 0 \quad .
\end{align}
We solve this system of equations by Fourier transformation in the two-dimensional coordinate space perpendicular to the interface, and then solving separately for the two half spaces with the boundary conditions stated previously in section \ref{chap:mestat}. We obtain the following expressions for the potentials $\phi_m$ and $\phi_e$ (for details see the Appendix): 
\begin{align}
\phi^{in}_e&=&\dfrac{c^{in}_{e1}}{\sqrt{R^2+|\zeta^+z-z_0|^2}}+\dfrac{c^{in}_{e2}}{\sqrt{R^2+|\zeta^-z-z_0|^2}}\label{eqn::e_in_potential}\\
\phi_e^{out}&=&\dfrac{1}{4\pi\epsilon_0}\dfrac{q}{\sqrt{R^2+|z-z_0|^2}}+\dfrac{c^{out}_{e1}}{\sqrt{R^2+|z+z_0|^2}}\\
\phi_m^{in} &=&\dfrac{c^{in}_{b1}}{\sqrt{R^2+|\zeta^-z-z_0|^2}}+\dfrac{c^{in}_{b2}}{\sqrt{R^2+|\zeta^+z-z_0|^2}}\label{eqn::m_in_potential}\\
\phi_m^{out}&=&\dfrac{c^{out}_{b1}}{\sqrt{R^2+|z+z_0|^2}}\label{eqn::mag_potential}
\end{align}
with $R=\sqrt{x^2+y^2}$. $\zeta^\pm=\sqrt{\frac{\pm\gamma+a+d}{2}}$ is determined by the electric, magnetic and magnetoelectric susceptibilities:
\begin{align*}
a&=\dfrac{\epsilon_\parallel\mu_\perp -\alpha_\parallel \alpha_\perp}{\epsilon_\parallel\mu_\parallel-\alpha_\parallel^2}\\
b&=\dfrac{\epsilon_\parallel \alpha_\perp-\epsilon_\perp\alpha_\parallel}{\epsilon_\parallel\mu_\parallel-\alpha_\parallel^2}\\
c&=\dfrac{\mu_\parallel\alpha_\perp-\mu_\perp\alpha_\parallel}{\epsilon_\parallel\mu_\parallel-\alpha_\parallel^2}\\
d&=\dfrac{\epsilon_\perp \mu_\parallel -\alpha_\parallel\alpha_\perp}{\epsilon_\parallel\mu_\parallel-\alpha_\parallel^2}\\
\gamma&=\sqrt{a^2-2 a d+4 b c+d^2}\quad .
\end{align*}
The values of the parameters for the case of Cr$_2$O$_3$ (obtained using the susceptibilities from table \ref{tab:parameters}) are given in table \ref{tab::val_cr2o3}.

Eqn.~\ref{eqn::mag_potential} leads us immediately to the central result of our calculations, which is that the magnetic field outside the material has the monopolar form:
\begin{equation}\label{eq::exact}
\vec{B}(\vec r)=\mu_0c^{out}_{b1}\dfrac{\vec r-\vec r_1}{|\vec r-\vec r_1|^{3}} \quad .
\end{equation}
Here $\vec r = (x,y,z)$ and $\vec r_1 =(0,0,-z_0)$. We plot the magnetic field in Fig. \ref{fig:plotfield} (c) for the parameters of Cr$_2$O$_3$. The monopolar nature above the surface is clear, while
the behavior beneath the surface is more complicated than in the isotropic case.

\begin{table}[h]
\begin{tabular}{|c|rlc|}
\hline 
$c_{e1}^{in}$ & & $2.48 \times 10^{-10}$ & Vm\\ 
\hline 
$c_{e2}^{in}$ & & $6.15 \times 10^{-16}$& Vm \\ 
\hline 
$c_{e1}^{out}$ & $-$&$7.47 \times 10^{-9}$ &Vm \\ 
\hline 
$c_{b1}^{in}$ & & $3.41 \times 10^{-15}$ &Am\\ 
\hline 
$c_{b2}^{in}$ & & $3.38 \times 10^{-15}$ &Am\\ 
\hline 
$c_{b1}^{out}$ & $-$ & $1.59 \times 10^{-16}$ &Am \\ 
\hline 
\end{tabular}
\caption{Calculated values for the coefficients in Eqns. (20) - (23), for an elementary point charge at a vacuum/Cr$_2$O$_3$ interface.\label{tab::val_cr2o3}}
\end{table}

Note that the electric field (not shown) is indistinguishable from that obtained for the isotropic case because it is dominated by the dielectric response, which is almost isotropic. The additional electric polarization that is induced by the magnetoelectric response is negligible compared to the direct dielectric response. We emphasize again that, due to the transformation properties of the magnetoelectric tensor, the sign of the magnetic image charges, and the corrsponding induced $\vec{B}$ field, will be opposite in the two different AFM domains of Cr$_2$O$_3$.

\subsection{Dependence of the monopolar field strength on the magnetoelectric anisotropy \label{sec:aniso_dep}}

\begin{figure}
\centering
\def\svgwidth{\columnwidth}
\includegraphics[width=\columnwidth]{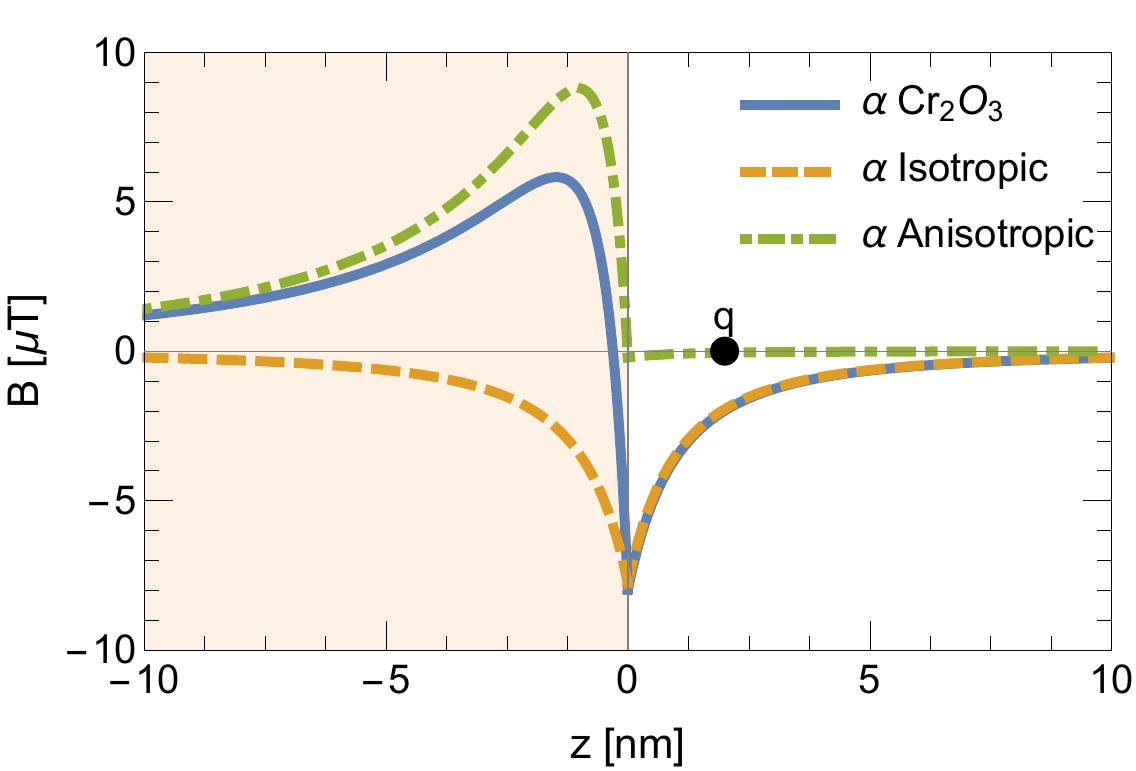}
\caption{Magnetic field $\vec{B}$ along $(0,0,z)$ induced by an electronic  charge , $q=+|e|$, 2 nm above the surface (at $z=0$) of the magnetoelectric slab. Positive $z$ values are above the sample surface. The field is decomposed into contributions from the sum and difference components of the magnetoelectric tensor.  We see that the monopolar field outside the sample is determined entirely by an isotropic component of the magnetoelectric response. \label{fig4}}
\end{figure}

We saw in the previous two sections that the induced monopolar field depends on both the magnitude of the magnetoelectric response and its anisotropy, that is the relative magnitudes of $\alpha_\parallel$ and $\alpha_\perp$. In Appendix C we give a detailed analysis of the effect of anisotropy, the main results of which we present here. In Fig.~\ref{fig4} we show the field contributions from the ``sum'' (proportional to the sum of $\alpha_\perp$ and $\alpha_\parallel$) and ``difference'' (proportional to the difference between $\alpha_\perp$ and $\alpha_\parallel$) components of the magnetoelectric tensor:
\begin{equation}
\mat{\alpha}=\frac{1}{2}(\alpha_\perp+\alpha_\|)\mathds{1}+\frac{1}{2}(\alpha_\perp-\alpha_\|)\begin{bmatrix} 1&0&0 \\ 0&1&0\\ 0 & 0 & -1\end{bmatrix} \quad ,
\end{equation}
calculated assuming that the anisotropies in $\epsilon$ and $\mu$ are small.

We see that, for this particular slab orientation (with the surface perpendicular to the high-symmetry axis), while both sum and difference components of the magnetoelectric tensor contribute to the field within the slab, only the sum component is relevant for the field outside the magnetoelectric; in fact for the case of exactly isotropic $\mat\epsilon$ and $\mat\mu$ tensors the field outside the slab is given by the result that we derived for the fully isotropic case, Eqn.~\eqref{eq::iso}
\begin{equation}\label{eq::iso}
\vec B(\vec r)=-\dfrac{\mu_0}{4\pi}\dfrac{q (\alpha_\perp+\alpha_\parallel)}{(\mu+\mu_0)(\epsilon+\epsilon_0)-\frac{1}{4}(\alpha_\perp+\alpha_\parallel)^2}\dfrac{\vec r-\vec r_1}{|\vec r - \vec r_1|^{3}}\quad .
\end{equation}
 This is consistent with the symmetry of the vacuum, in which a hypothetical magnetic charge would induce a purely monopolar magnetic field. Anisotropies in the $\mat\epsilon$ and $\mat\mu$ tensors modify the magnitude of $B(r)$ slightly from that of Eqn.~\eqref{eq::iso} (for the case of Cr$_2$O$_3$ using the values from table \ref{tab:parameters} we find a difference of 0.05\% between the exact solution and that for averaged isotropic $\mat\epsilon$ and $\mat\mu$), but do not change its monopolar form. 

\begin{figure}
\includegraphics[width=\columnwidth]{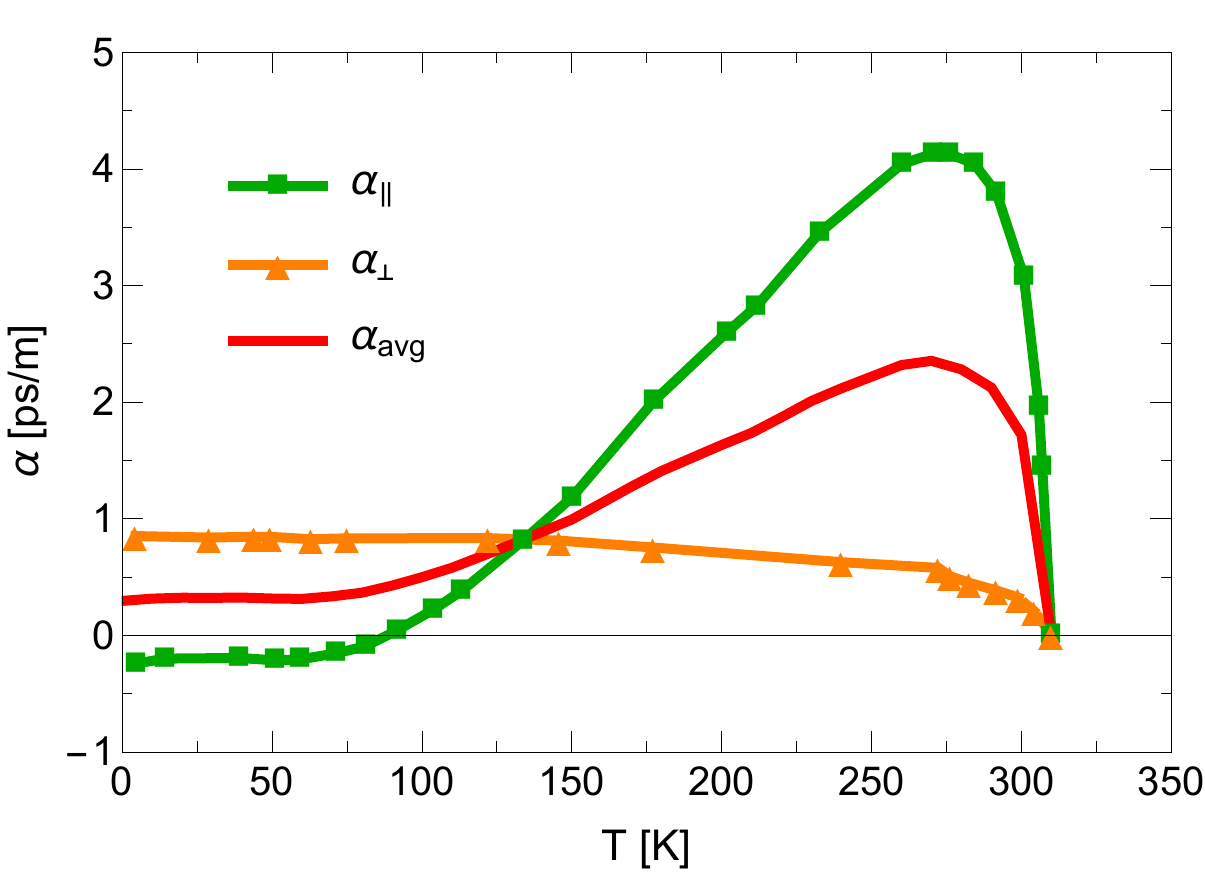}
\caption{Measured temperature dependence of the parallel ($\alpha_{\parallel}$, green squares) and perpendicular ($\alpha_{\perp}$, orange triangles) magnetoelectric response in Cr$_2$O$_3$. The red circles show the average, $\frac{1}{2}(\alpha_\parallel+\alpha_\perp)$. Data taken from Ref~\onlinecite{Wiegelmann:1994tq}. }\label{fig:alphat}
\end{figure}

This feature makes it particularly straightforward to predict the temperature dependence of the monopolar field. The highly temperature dependent magnetoelectric response in Cr$_2$O$_3$\cite{Wiegelmann:1994tq} is reproduced in Fig.~\ref{fig:alphat}. While the in-plane magnetoelectric response, $\alpha_\perp$, shows the usual Brillouin-function form below the N\'eel temperature (orange triangles in Fig.~\ref{fig:alphat}), the spin-fluctuation mechanism\cite{Mostovoy:2010ia} responsible for the out-of-plane response, $\alpha_\parallel$, results in a strong temperature dependence (green squares in Fig.~\ref{fig:alphat}), with $\alpha_\parallel$ even changing sign at low temperature. Since the strength of the induced monopole is proportional to the sum, $\frac{1}{2}(\alpha_\parallel+\alpha_\perp)$, shown as the red line in Fig. \ref{fig:alphat}, the corresponding induced monopolar field must have the same temperature dependence. We see that the induced monopolar field should increase with increasing temperature, 
reaching a maximum at around 280~K, before decreasing and vanishing at the N\'eel temperature at $\sim 310$ K.

\section{Experimental search for the magnetic monopole using Low energy muon spin rotation (LE-$\mu$SR)}\label{sec:muons}

Next we describe our experimental search for the magnetic monopolar field using low energy muon spin rotation \cite{Morenzoni1994prl,Prokscha2008nima,morenzoni-2000,morenzoni-2004}  (LE-$\mu$SR). 

\subsection{Experimental setup}

In the LE-$\mu$SR method, fully polarized muons are implanted into a sample and the local magnetic field at the muon stopping site is measured by monitoring the evolution of the muon spin polarization. This is achieved via the anisotropic beta decay positron which is emitted preferentially in the direction of the muon's spin at the time of decay. Using appropriately positioned detectors one can measure the asymmetry, $A(t)$, of the beta decay along the initial polarization direction. $A(t)$ is proportional to the time evolution of the spin polarization of the ensemble of implanted spin probes~\cite{Yaouanc2010}. 

Conventional $\mu$SR experiments use so-called surface muons with an implantation energy of \SI{4.1}{MeV}, resulting in a
stopping range in typical density solids of from \SI{0.1}{mm} to \SI{1}{mm} below the surface. As a result, their application is limited to studies of bulk properties and they cannot provide depth-resolved information or study extremely thin film samples. In contrast, depth-resolved $\mu$SR measurements can be performed at the low-energy muon (LEM) spectrometer using muons with tunable kinetic energies in the \SIrange{1}{30}{\keV} range, corresponding to implantation depths of \SIrange{10}{200}{\nm}. We take advantage of this capability here.

\begin{figure}
\centering
\includegraphics[width=\columnwidth]{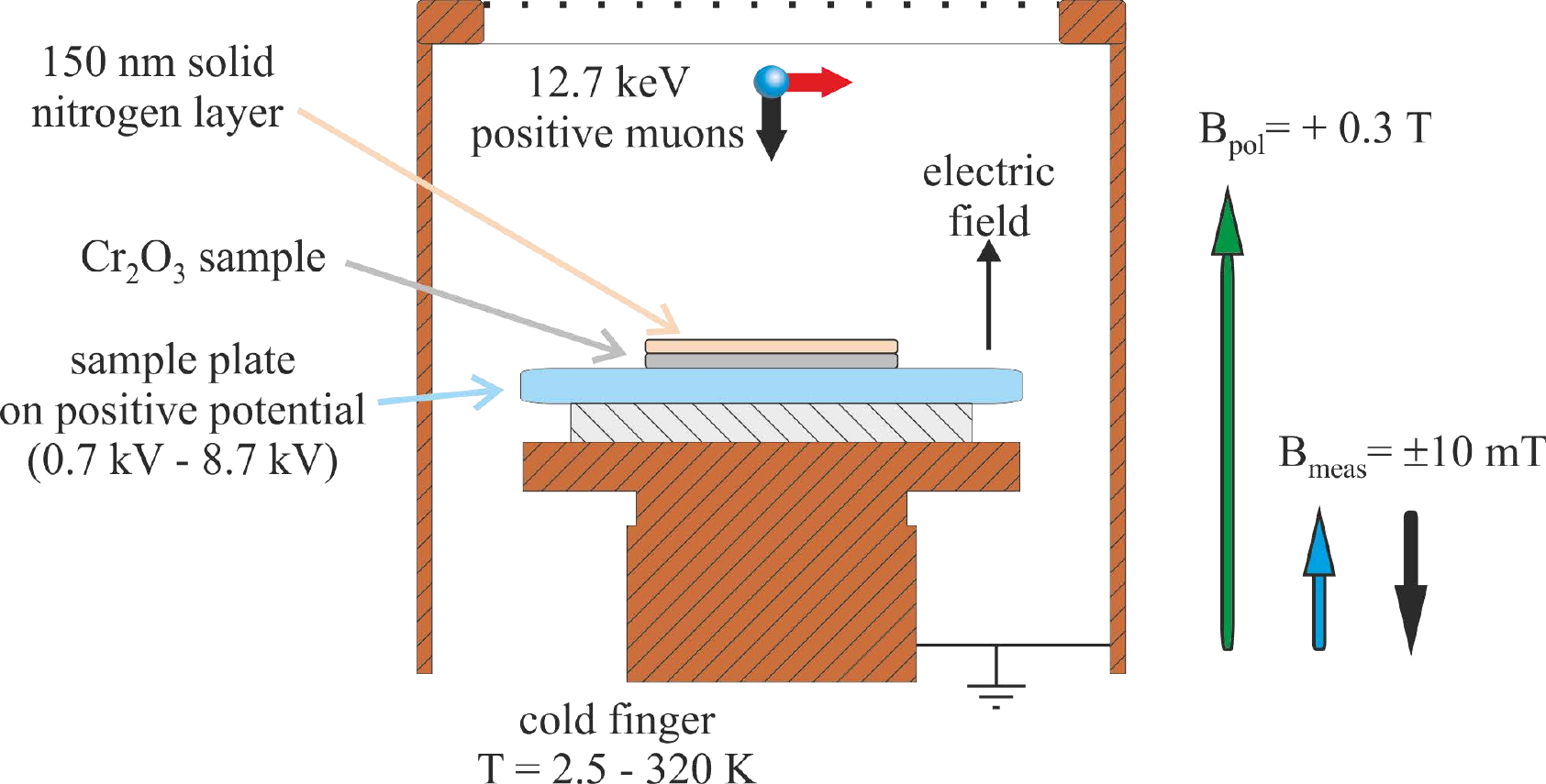}
\caption{Sketch of the LE-$\mu$SR setup used for this experiment. Muons with a kinetic energy of 12.7 keV enter the sample region with nearly 100\% spin polarization (red arrow). The energy of the muons impinging on the sample can be tuned by choosing the appropriate potential at the sample plate. The sample was cooled in a positive poling field of $\vec{B}_{\rm pol}=0.3$~T and a positive electric field of $\vec{E} > 1$ kV/cm to ensure a positive $\alpha$. The measurements at low temperature were performed in $\vec{B}_{\rm meas}= \pm 10$~mT.\label{fig:muon_setup}}
\end{figure}

\begin{figure}
\centering
\begin{flushleft}
\includegraphics[width=0.95\columnwidth]{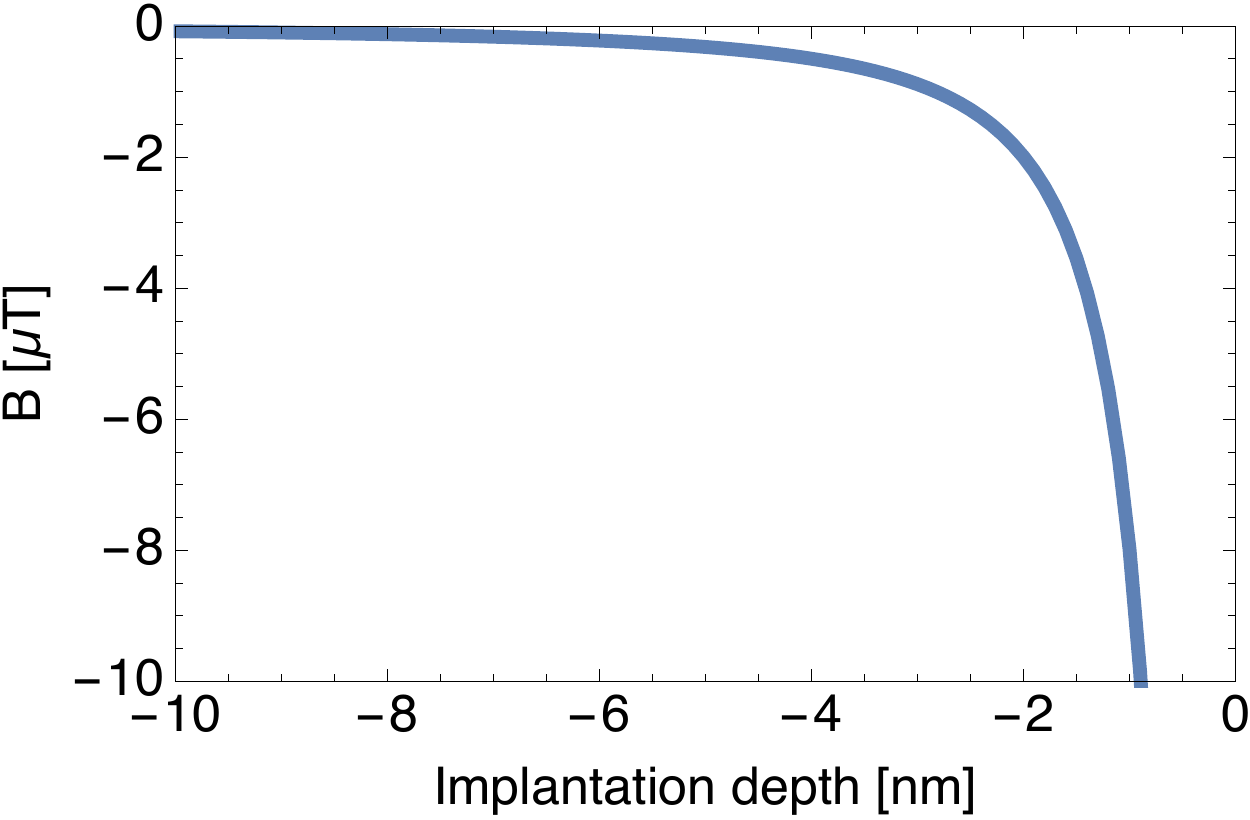}\\
\end{flushleft}
\caption{Calculated magnetic field at the site of a muon at a distance of minus the implantation depth above the surface of Cr$_2$O$_3$.
\label{fig:muon_fraction} }
\end{figure}

Our measurement, which builds on our previous attempt to measure the image monopole in topological insulators\cite{Salman_et_al:2018}, is designed in the following way: We use a 500 nm thick Cr$_2$O$_3$ film grown in the (001) direction, which is coated by an insulating stopping layer, in this case solid nitrogen, N$_2$. The muons (which carry a positive electronic charge $+e$) are implanted at different depths in the N$_2$ layer. The electric field of the muon should penetrate into the Cr$_2$O$_3$ layer and induce both electric and magnetic responses, with the magnetic response being the monopolar field described above in Section~\ref{Section2}. The muon itself then acts as the magnetic probe to measure the induced magnetic field. The full experimental setup is sketched in Fig.~\ref{fig:muon_setup}. In Fig.~\ref{fig:muon_fraction} we show the calculated magnetic field as a function of the distance of the muon from the Cr$_2$O$_3$ surface, and note that, like the field from a charge at a fixed point shown in Fig.~\ref{fig4}, it has a $\frac{1}{d^2}$ dependence.

The Cr$_2$O$_3$ films used here were grown by reactive rf sputtering on (0001) Al$_2$O$_3$ substrates using a metal Cr target in an Ar + O$_2$ atmosphere (base pressure $< 1 \times 10^{-6}$ Pa) at a substrate temperature of 773 K. Bottom Pt electrodes with thicknesses of 25 nm were sputtered on Al$_2$O$_3$ substrates and Cr$_2$O$_3$ films using shadow masking. Prior to our measurements, the Cr$_2$O$_3$ layer was prepared in a single domain state using magnetoelectric annealing. This was achieved by cooling the sample from 320~K through the N\'eel temperature to 20~K in a positive magnetic field of 0.3~T and a positive electric field larger than 1~kV/cm, both applied along the surface normal. Since $E$ and $H$ are parallel, such an anneal yields a single magnetoelectric domain with positive magnetoelectric tensor $\alpha$ \cite{mahdawi-2017}. 

We then deposited a 150 nm layer of solid nitrogen on top of the Cr$_2$O$_3$ film to provide an insulating muon stopping region above the surface of the magnetoelectric. The N$_2$ deposition and all subsequent measurements were performed at 20K to maintain the N$_2$ in the solid state.
Muons were then implanted into this bilayer structure with different incident muon kinetic energies, in the presence of a small bias field, $B_{\rm meas}= \pm 10$~mT. The fraction of muons that do not capture an electron to form the neutral hydrogen-like muonium state is about 40\% in the N$_2$ film \cite{PhysRevLett.98.227401}. The muonium response occurs at a completely different resonance frequency and so is easily subtracted from the measurement.
The bias field is used to increase the accuracy of the measurement, but is too small to reorient the antiferromagnetic domain  and so does not change the sign of the magnetoelectric tensor \cite{Borisov2016}. We performed independent second harmonic generation domain imaging experiments\cite{Fiebig:2005} and verified that the domain structure is stable up to fields of 5.8~T (at which a spin-flop occurs).

\begin{figure}
\centering
\includegraphics[width=0.95\columnwidth]{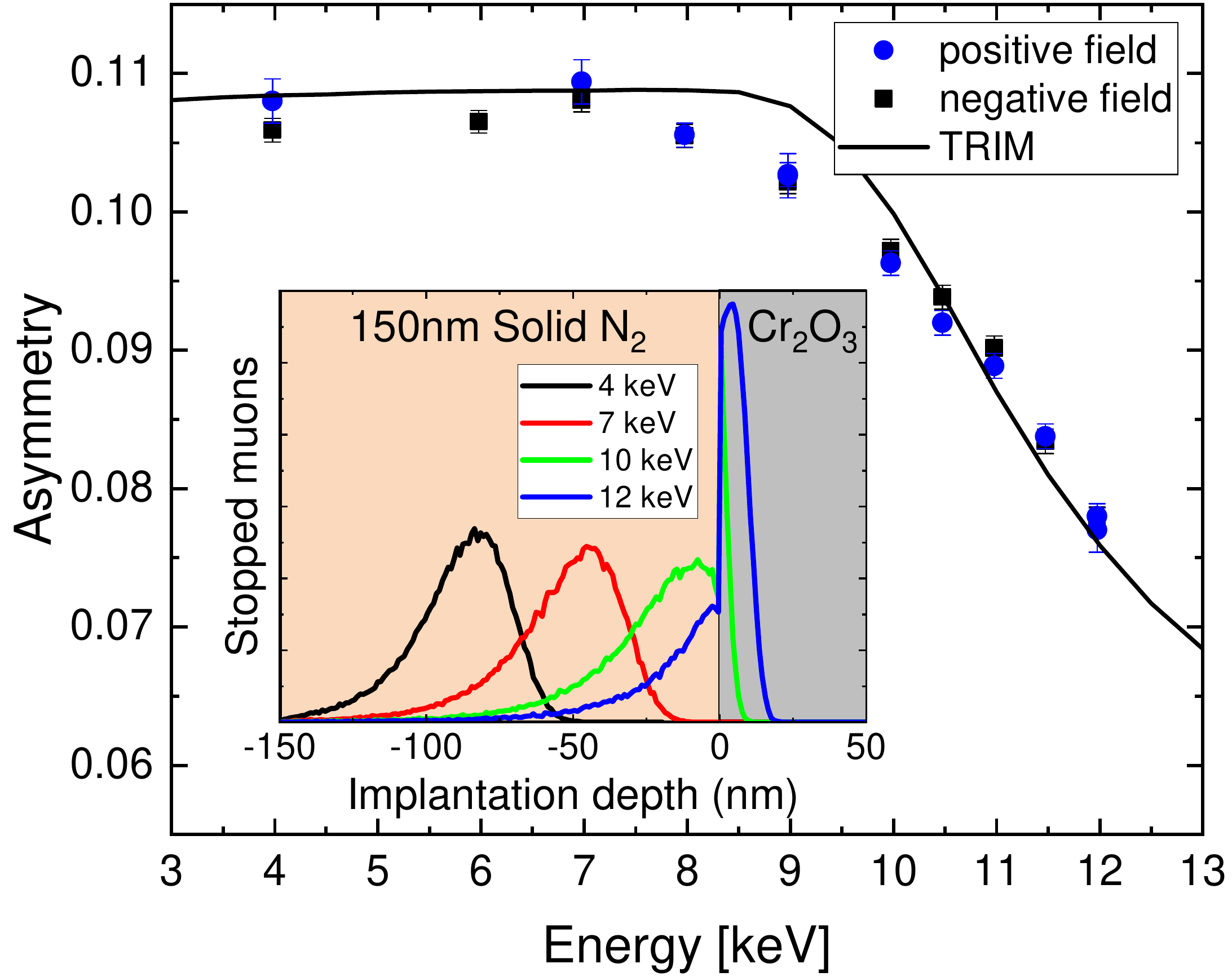}\caption{Measured LE-$\mu$SR asymmetry (blue circles and black squares) for positive and negative applied magnetic fields and TRIM.SP prediction (solid black line) as a function of muon implantation energy  for a 150nm thick solid nitrogen layer on Cr$_2$O$_3$. Inset: Muon stopping profiles calculated using TRIM.SP for various muon implantation energies.\label{fig:muon_stopping} }
\end{figure}

In Fig.~\ref{fig:muon_stopping} (inset) we show the muon stopping profiles (that is the fraction of muons as a function of implantation depth) for different muon implantation energies, calculated assuming an N$_2$ thickness of 150 nm and a density of 1 g/cm$^3$. We used the Monte Carlo program TRIM.SP, which treats the positive muon as a light proton and has been shown to be accurate for low-energy muons \cite{Morenzoni2002nimb}.  The black line in the main panel shows the calculated LE-$\mu$SR initial asymmetry as a function of implantation energy calculated from these stopping profiles, with the assumption that only those muons that stop in the N$_2$ layer and that do not form muonium contribute to the initial polarization.  The initial asymmetry decreases for increasing implantation energies as the muons enter the magnetic Cr$_2$O$_3$ layer where they quickly lose their polarization due to the strong internal magnetic fields. Also plotted in the main panel is our observed LE-$\mu$SR asymmetry measured in a transverse magnetic field of $\pm$10~mT. The agreement in trend between the results based on the TRIM.SP calculations and the measured values indicate that our assumed values for the thickness and density of the nitrogen layer are reasonable.

\subsection{Results}\label{sec:muonresults}

\begin{figure}
\centering
\includegraphics[width=\columnwidth]{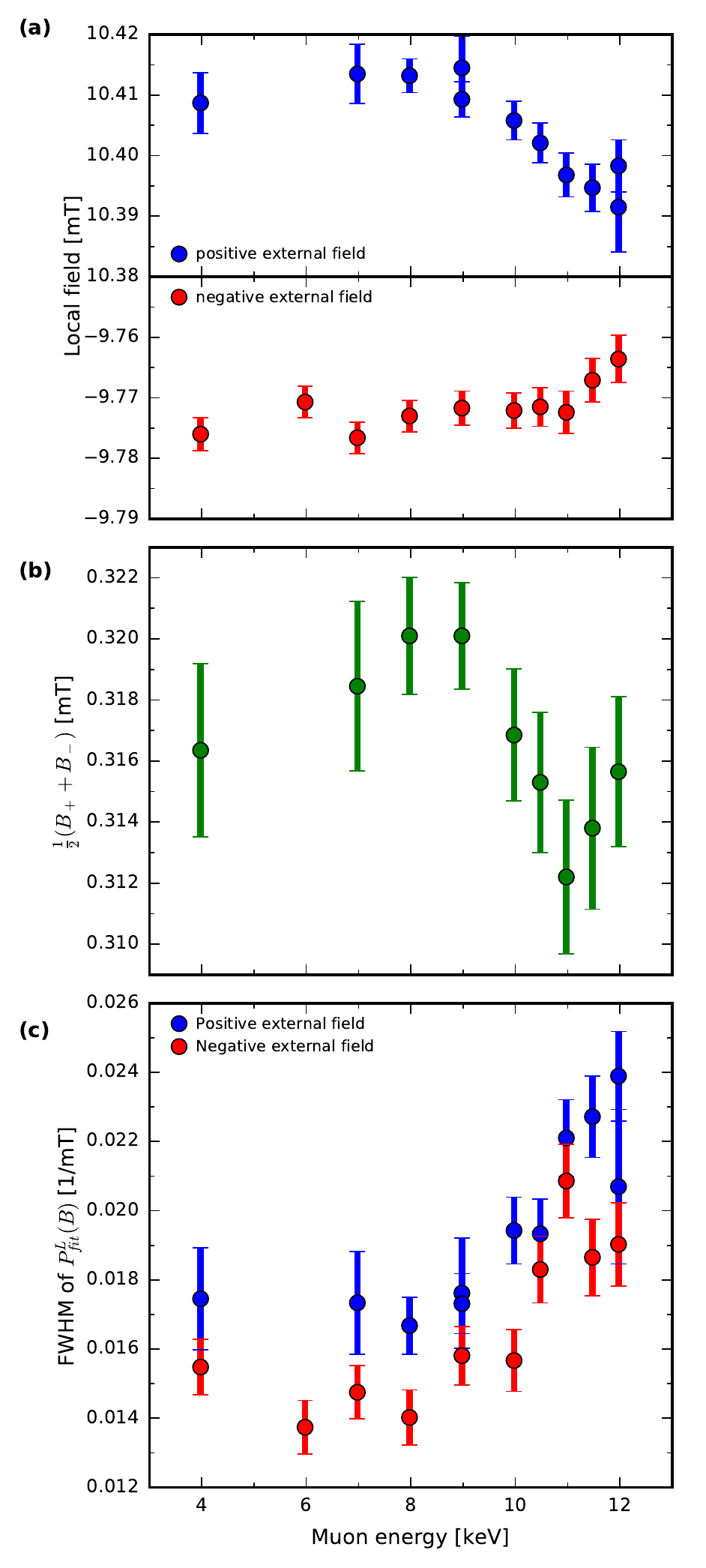}
\caption{(a) Measured magnetic field experienced by the muons stopping in the solid nitrogen layer as a function of energy. With increasing implantation energy the average distance of muons to the Cr$_2$O$_3$ interface decreases. Positive and negative external field refers to the small bias field applied at low temperature for the duration of the measurement; in both cases the sample was poled in a large positive field parallel to the external electric field prior to the measurement, to prepare it in a single magnetoelectric domain with positive $\alpha$. (b) Measured local field corrected for contributions from parasitic field effects as a function of muon implantation energy. (c) Measured full width at half maximum (FWHM) of the magnetic field distribution fit to a Lorentzian distribution, $P^\textrm{L}_\textrm{fit}(B)$, as a function of muon implantation energy. \label{fig:muon_FWHM} }
\end{figure}

In Fig.~\ref{fig:muon_FWHM} (a), we show the measured internal fields at the muon sites as a function of the muon implantation energy, with higher implantation energies corresponding to smaller average distances to the Cr$_2$O$_3$ surface. The upper panel (blue circles) shows the results obtained in small positive bias field (along $+c$), and the lower panel (red circles) those obtained in a small negative bias field. The local field shown in  Fig. \ref{fig:muon_FWHM} is the sum of the bias field plus any internal field at the muon site. We see that in both cases the muon experiences a local magnetic field that varies monotonically with its distance from the surface. Note again that only the muons stopping in the nitrogen overlayer contribute to the signal as the muons stopping in Cr$_2$O$_3$ quickly depolarize. The LE-$\mu$SR raw-data for an example point is shown in Appendix \ref{app:musr}.

If the only contribution to the internal field at the muon site were the monopolar field from the magnetoelectric response, we would expect the shifts in both cases to be in the same direction, since both sets of measurements are performed on the same magnetoelectric domain. It is known, however, that Cr$_2$O$_3$ thin films can have stray spins caused by defects at the interface with the Al$_2$O$_3$ substrate, as well as impurity spins at the interface, which have been shown to be susceptible to small magnetic fields in thin film samples\cite{Borisov2016,Appel_et_al:2018}. (Note that the intrinsic surface spin density resulting from the termination of the antiferromagnetic magnetoelectric\cite{Andreev1996,Astrov1996,fallarino} is not reversed under the conditions of our experiment, since this would require the reversal of the full antiferromagnetic domain\cite{fallarino}. In addition, it has an associated field that is negligible compared with that from the monopole effect.) To remove the contribution from the stray magnetic dipoles, which we expect to switch with the applied magnetic bias field, we therefore sum the local internal values obtained in positive and negative bias, and present $0.5(\textit{B$_+$}+\textit{B$_-$})$ as a function of muon energy in Fig. \ref{fig:muon_FWHM} (b). The base level bias corresponds to the switching precision of the small magnetic bias field.
We obtain an internal field shift that is consistent with the expected behavior of the induced magnetic monopole: The maximum value close to the surface is of the same order of magnitude (several $\mu$T) as the calculated value, the sign is as expected  for the prepared magnetoelectric domain, and it decays with distance from the interface. While the size of the error bars prohibits extraction of the exact functional form, the decay is consistent with quadratic behaviour. 

For completeness, we present in Fig.~\ref{fig:muon_FWHM} (c) the full width at half maximum of the distribution of the fields sensed by the muons, $P^\textrm{L}_\textrm{fit}(B)$, extracted from the damping rate of the measured muon spin polarization assuming a Lorentzian field distribution. The values of the linewidths of $\sim$20$\mu$T, and their increase towards the Cr$_2$O$_3$ surface, are consistent with the total local field values from panel (a) convoluted with the calculated stopping profiles shown earlier, indicating that this incoherent broadening results primarily from the contribution from the stray spins that are aligned ferromagnetically by the bias field.

\section{Discussion and other experimental techniques}

The small field shift in our LE-$\mu$SR measurements, combined with the increased width of the field distribution towards the interface present a first hint that a monopole is indeed induced by an electric charge at a magnetoelectric surface. In this final section we discuss studies that we have attempted using other techniques, as well as additional possible future routes for confirmation of the monopole's existence. 

A first step would be to perform temperature-dependent measurements using the LE-$\mu$SR technique described above. We showed in section \ref{sec:aniso_dep} the temperature dependence of the average magnetoelectric response, which in turn determines the strength of the monopolar field. A measured increase in field strength on warming with a maximum at around 280~K would be a strong indication that the origin of the field is the magnetoelectric response of the sample. For such a study, a different stopping layer would be needed because nitrogen would not be solid.

\subsection{Magnetic force microscopy}

\begin{figure}
\includegraphics[width=\columnwidth]{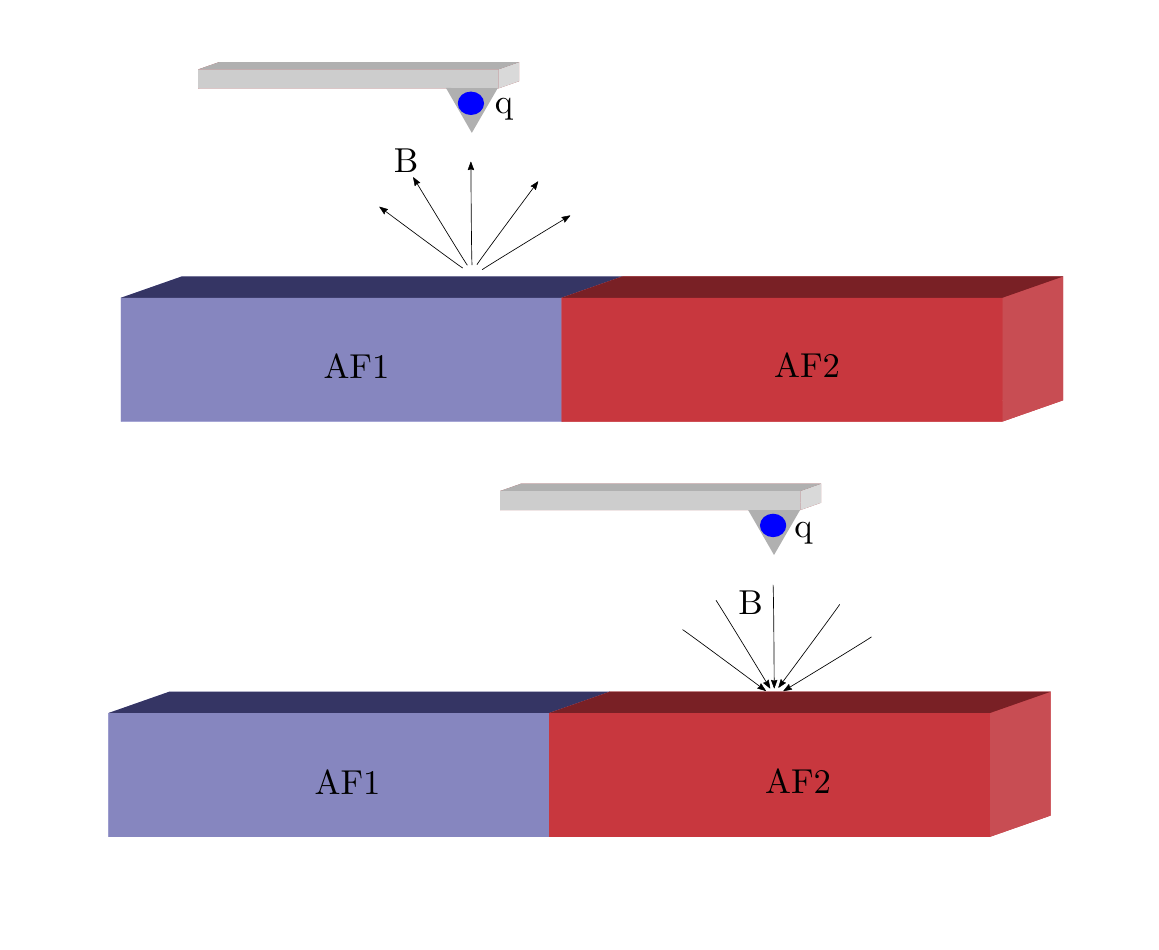}
\caption{Proposed technique for measuring monopolar magnetic fields. The charged MFM tip both induces the image monopole and detects its field. AF1 and AF2 indicate oppositely oriented antiferromagnetic domains, which support monopoles and associated fields of opposite sign.}\label{fig:MFM}
\end{figure}

In addition to the muon experiments we performed magnetic force microscopy (MFM) on a cut and etch-polished commercial c-oriented Cr$_2$O$_3$ crystal of $d$ = 150 $\mu$m thickness grown by the Verneuil method (Kristallhandel Kelpin). The magnetic tip of an atomic force microscope acted as an electric charge monopole by applying a voltage $U$ of 20 V between the tip and the copper back electrode of the sample. At the same time, the magnetization of the tip served as the detector for the induced monopolar magnetic field. 
The goal of the experiment was to exploit the different sign of $\alpha$ for the two antiferromagnetic domains and measure a change of sign in the response when the tip moves across a domain boundary, as sketched in Fig.~\ref{fig:MFM}. In addition we aimed to vary the tip-surface distance to verify the characteristic $r^2$ dependence of a monopolar field. From our values of $U$, $d$ and $\alpha$ we estimated the monopolar field at the position of the tip to be on the order of 1 $\mu T$ which should be detectable as a change of the mechanical deformation of the magnetized tip. 

In the first step, we determined the distribution of antiferromagnetic domains in our Cr$_2$O$_3$ samples by optical second harmonic generation \cite{Fiebig:2005}. In step two, we corroborated the sensitivity of our experiment to the magnetization induced via the linear magnetoelectric effect. We coated a Cr$_2$O$_3$ sample with a metallic platinum film of 50 nm thickness acting as front electrode and detected the Cr$_2$O$_3$ bulk magnetization induced by 50V applied to the electrodes. This revealed a domain-dependent magnetization one to two orders above our detection limit \cite{Schoenherr_et_al:2017}. In the third step, we repeated the experiment on an uncoated Cr$_2$O$_3$ sample, now employing the charged tip as the source of charge to generate a monopolar magnetic field as described above. We found, however, that the residual Cr$_2$O$_3$ surface roughness of about 4 nm led to a pronounced electrostatic inhomogeneity in this insulating sample that obscured any response expected from the magnetic-monopole field. 
No signal difference was detected at the position of the antiferromagnetic domain boundaries.

\subsection{Scanning SQUID magnetometry}

Another possible technique for measuring the induced monopolar field could be scanning SQUID magnetometry. When a charge $n_e \times |q|$ is placed on the magnetoelectric surface, we have seen that the resulting monopole is given by 
\begin{eqnarray}
m &\approx & -\dfrac{\mu_0}{4\pi}\dfrac{q(\alpha_\perp+\alpha_\parallel)}{(\epsilon+\epsilon_0)(\mu+\mu_0)-\frac{1}{4}(\alpha_\perp+\alpha_\parallel)^2} \\
& = & n_{e}\cdot 1.92\cdot10^{-22} \text{Tm}^2
\end{eqnarray}
for the case of Cr$_2$O$_3$. 
The magnetic flux from the magnetic monopole through a Josephson junction is then given by (see derivation in Appendix)
\begin{align}
\Phi&=\int \textbf{B}\cdot d\textbf{S}=m(z+d)\int\limits_{0}^{2\pi}d\phi\int\limits_0^Rrdr \dfrac{1}{(r^2+(z+d)^2)^{3/2}}\\
&=2\pi m\left(1-\dfrac{z+d}{\sqrt{R^2+(z+d)^2}}\right) \quad ,
\end{align}
where $m$ is the magnetic monopole moment, $z$ is the distance of the pickup from the interface, $d$ is the distance of the charge from the interface, and $R$ is the radius of the loop. One of the key challenges in this experiment would be to find a way to fix and localize charge above the surface.

\section{Conclusions}
In summary, we derived the form of the electric and magnetic fields that are induced by an electric charge above a surface of a semi-infinite slab of magnetoelectric material. We found that, for both isotropic and uniaxial magnetoelectrics, the electric charge induces a magnetic image charge, which is the source of a monopolar field decaying with $r^{2}$ in the vacuum region. The strength of this induced field depends on the value of the sum part $\dfrac{1}{2}|\alpha_\perp+\alpha_\parallel|$ of the magnetoelectric tensor and any internal field arising from a difference component of the magnetoelectric response vanishes at the interface. We showed that the magnitude of the response induced by a single electronic charge is large enough to be detectable experimentally, and described searches using muon spin spectroscopy and magnetic force microscopy. Our muon spin spectroscopy data, while not fully conclusive, are consistent with the existence of the monopolar field. We hope that our encouraging initial results, as well as our discussion of other possible experimental approaches for measurement of the monopole, motivate further studies.

\section{Acknowledgements}
The low energy $\mu$SR experiments were performed at the Swiss Muon Source S$\mu$S, Paul Scherrer Institute, Switzerland.
Work at ETH was supported financially by the ETH Zurich, by the ERC Advanced Grant program, No. 291151, by the Max R\"ossler Prize of the ETH, and by the Sinergia program of the Swiss National Science Foundation Grant No. CRSII2\_147606/1. 
This work was partly funded by ImPACT Program of Council for Science, Technology and Innovation (Cabinet Office, Japan Government).

\bibliography{bib}

\begin{thebibliography}{34}%
\makeatletter
\providecommand \@ifxundefined [1]{%
 \@ifx{#1\undefined}
}%
\providecommand \@ifnum [1]{%
 \ifnum #1\expandafter \@firstoftwo
 \else \expandafter \@secondoftwo
 \fi
}%
\providecommand \@ifx [1]{%
 \ifx #1\expandafter \@firstoftwo
 \else \expandafter \@secondoftwo
 \fi
}%
\providecommand \natexlab [1]{#1}%
\providecommand \enquote  [1]{``#1''}%
\providecommand \bibnamefont  [1]{#1}%
\providecommand \bibfnamefont [1]{#1}%
\providecommand \citenamefont [1]{#1}%
\providecommand \href@noop [0]{\@secondoftwo}%
\providecommand \href [0]{\begingroup \@sanitize@url \@href}%
\providecommand \@href[1]{\@@startlink{#1}\@@href}%
\providecommand \@@href[1]{\endgroup#1\@@endlink}%
\providecommand \@sanitize@url [0]{\catcode `\\12\catcode `\$12\catcode
  `\&12\catcode `\#12\catcode `\^12\catcode `\_12\catcode `\%12\relax}%
\providecommand \@@startlink[1]{}%
\providecommand \@@endlink[0]{}%
\providecommand \url  [0]{\begingroup\@sanitize@url \@url }%
\providecommand \@url [1]{\endgroup\@href {#1}{\urlprefix }}%
\providecommand \urlprefix  [0]{URL }%
\providecommand \Eprint [0]{\href }%
\providecommand \doibase [0]{http://dx.doi.org/}%
\providecommand \selectlanguage [0]{\@gobble}%
\providecommand \bibinfo  [0]{\@secondoftwo}%
\providecommand \bibfield  [0]{\@secondoftwo}%
\providecommand \translation [1]{[#1]}%
\providecommand \BibitemOpen [0]{}%
\providecommand \bibitemStop [0]{}%
\providecommand \bibitemNoStop [0]{.\EOS\space}%
\providecommand \EOS [0]{\spacefactor3000\relax}%
\providecommand \BibitemShut  [1]{\csname bibitem#1\endcsname}%
\let\auto@bib@innerbib\@empty
\bibitem [{\citenamefont {Dirac}(1931)}]{Dirac:1931vc}%
  \BibitemOpen
  \bibfield  {author} {\bibinfo {author} {\bibfnamefont {P.A.M.}\ \bibnamefont
  {Dirac}},\ }\bibfield  {title} {\enquote {\bibinfo {title} {{Quantised
  singularities in the electromagnetic field}},}\ }\href@noop {} {\bibfield
  {journal} {\bibinfo  {journal} {Proc. R. Soc. London, Ser. A}\ }\textbf
  {\bibinfo {volume} {133}},\ \bibinfo {pages} {60--72} (\bibinfo {year}
  {1931})}\BibitemShut {NoStop}%
\bibitem [{\citenamefont {Rajantie}(2016)}]{Rajantie:2016}%
  \BibitemOpen
  \bibfield  {author} {\bibinfo {author} {\bibfnamefont {A.}~\bibnamefont
  {Rajantie}},\ }\bibfield  {title} {\enquote {\bibinfo {title} {{The search
  for magnetic monopoles}},}\ }\href@noop {} {\bibfield  {journal} {\bibinfo
  {journal} {Phys. Today}\ }\textbf {\bibinfo {volume} {69}},\ \bibinfo {pages}
  {40--46} (\bibinfo {year} {2016})}\BibitemShut {NoStop}%
\bibitem [{\citenamefont {Castelnovo}\ \emph {et~al.}(2008)\citenamefont
  {Castelnovo}, \citenamefont {Moessner},\ and\ \citenamefont
  {Sondhi}}]{Castelnovo:2008hb}%
  \BibitemOpen
  \bibfield  {author} {\bibinfo {author} {\bibfnamefont {C.}~\bibnamefont
  {Castelnovo}}, \bibinfo {author} {\bibfnamefont {R.}~\bibnamefont
  {Moessner}}, \ and\ \bibinfo {author} {\bibfnamefont {S.~L.}\ \bibnamefont
  {Sondhi}},\ }\bibfield  {title} {\enquote {\bibinfo {title} {{Magnetic
  monopoles in spin ice}},}\ }\href@noop {} {\bibfield  {journal} {\bibinfo
  {journal} {Nature}\ }\textbf {\bibinfo {volume} {451}},\ \bibinfo {pages}
  {42--45} (\bibinfo {year} {2008})}\BibitemShut {NoStop}%
\bibitem [{\citenamefont {Morris}\ \emph {et~al.}(2009)\citenamefont {Morris},
  \citenamefont {Tennant}, \citenamefont {Grigera}, \citenamefont {Klemke},
  \citenamefont {Castelnovo}, \citenamefont {Moessner}, \citenamefont
  {Czternasty}, \citenamefont {Meissner}, \citenamefont {Rule}, \citenamefont
  {Hoffmann}, \citenamefont {Kiefer}, \citenamefont {Gerischer}, \citenamefont
  {Slobinsky},\ and\ \citenamefont {Perry}}]{Morris:2009kh}%
  \BibitemOpen
  \bibfield  {author} {\bibinfo {author} {\bibfnamefont {D.~J.~P.}\
  \bibnamefont {Morris}}, \bibinfo {author} {\bibfnamefont {D.~A.}\
  \bibnamefont {Tennant}}, \bibinfo {author} {\bibfnamefont {S.~A.}\
  \bibnamefont {Grigera}}, \bibinfo {author} {\bibfnamefont {B.}~\bibnamefont
  {Klemke}}, \bibinfo {author} {\bibfnamefont {C.}~\bibnamefont {Castelnovo}},
  \bibinfo {author} {\bibfnamefont {R.}~\bibnamefont {Moessner}}, \bibinfo
  {author} {\bibfnamefont {C.}~\bibnamefont {Czternasty}}, \bibinfo {author}
  {\bibfnamefont {M.}~\bibnamefont {Meissner}}, \bibinfo {author}
  {\bibfnamefont {K.~C.}\ \bibnamefont {Rule}}, \bibinfo {author}
  {\bibfnamefont {J.~U.}\ \bibnamefont {Hoffmann}}, \bibinfo {author}
  {\bibfnamefont {K.}~\bibnamefont {Kiefer}}, \bibinfo {author} {\bibfnamefont
  {S.}~\bibnamefont {Gerischer}}, \bibinfo {author} {\bibfnamefont
  {D.}~\bibnamefont {Slobinsky}}, \ and\ \bibinfo {author} {\bibfnamefont
  {R.~S.}\ \bibnamefont {Perry}},\ }\bibfield  {title} {\enquote {\bibinfo
  {title} {{Dirac Strings and Magnetic Monopoles in the Spin Ice
  Dy$_2$Ti$_2$O$_7$}},}\ }\href@noop {} {\bibfield  {journal} {\bibinfo
  {journal} {Science}\ }\textbf {\bibinfo {volume} {326}},\ \bibinfo {pages}
  {411--414} (\bibinfo {year} {2009})}\BibitemShut {NoStop}%
\bibitem [{\citenamefont {Khomskii}(2014)}]{Khomskii:2014dp}%
  \BibitemOpen
  \bibfield  {author} {\bibinfo {author} {\bibfnamefont {D.~I.}\ \bibnamefont
  {Khomskii}},\ }\bibfield  {title} {\enquote {\bibinfo {title} {{Magnetic
  monopoles and unusual dynamics of magnetoelectrics}},}\ }\href@noop {}
  {\bibfield  {journal} {\bibinfo  {journal} {Nat. Commun.}\ }\textbf {\bibinfo
  {volume} {5}},\ \bibinfo {pages} {4793} (\bibinfo {year} {2014})}\BibitemShut
  {NoStop}%
\bibitem [{\citenamefont {Fechner}\ \emph {et~al.}(2014)\citenamefont
  {Fechner}, \citenamefont {Spaldin},\ and\ \citenamefont
  {Dzyaloshinskii}}]{Fechner:2014us}%
  \BibitemOpen
  \bibfield  {author} {\bibinfo {author} {\bibfnamefont {M.}~\bibnamefont
  {Fechner}}, \bibinfo {author} {\bibfnamefont {N.~A.}\ \bibnamefont
  {Spaldin}}, \ and\ \bibinfo {author} {\bibfnamefont {I.~E.}\ \bibnamefont
  {Dzyaloshinskii}},\ }\bibfield  {title} {\enquote {\bibinfo {title} {{The
  magnetic field generated by a charge in a uniaxial magnetoelectric
  material}},}\ }\href@noop {} {\bibfield  {journal} {\bibinfo  {journal}
  {Phys. Rev. B}\ }\textbf {\bibinfo {volume} {89}},\ \bibinfo {pages} {184415}
  (\bibinfo {year} {2014})}\BibitemShut {NoStop}%
\bibitem [{\citenamefont {Qi}\ \emph {et~al.}(2008)\citenamefont {Qi},
  \citenamefont {Hughes},\ and\ \citenamefont {Zhang}}]{Qi:2008eu}%
  \BibitemOpen
  \bibfield  {author} {\bibinfo {author} {\bibfnamefont {X.-L.}\ \bibnamefont
  {Qi}}, \bibinfo {author} {\bibfnamefont {T.~L.}\ \bibnamefont {Hughes}}, \
  and\ \bibinfo {author} {\bibfnamefont {S.-C.}\ \bibnamefont {Zhang}},\
  }\bibfield  {title} {\enquote {\bibinfo {title} {{Topological field theory of
  time-reversal invariant insulators}},}\ }\href@noop {} {\bibfield  {journal}
  {\bibinfo  {journal} {Phys. Rev. B}\ }\textbf {\bibinfo {volume} {78}},\
  \bibinfo {pages} {195424} (\bibinfo {year} {2008})}\BibitemShut {NoStop}%
\bibitem [{\citenamefont {Qi}\ \emph {et~al.}(2009)\citenamefont {Qi},
  \citenamefont {Li}, \citenamefont {Zang},\ and\ \citenamefont
  {Zhang}}]{Qi:2009ip}%
  \BibitemOpen
  \bibfield  {author} {\bibinfo {author} {\bibfnamefont {X.-L.}\ \bibnamefont
  {Qi}}, \bibinfo {author} {\bibfnamefont {R.}~\bibnamefont {Li}}, \bibinfo
  {author} {\bibfnamefont {J.}~\bibnamefont {Zang}}, \ and\ \bibinfo {author}
  {\bibfnamefont {S.-C.}\ \bibnamefont {Zhang}},\ }\bibfield  {title} {\enquote
  {\bibinfo {title} {{Inducing a Magnetic Monopole with Topological Surface
  States}},}\ }\href@noop {} {\bibfield  {journal} {\bibinfo  {journal}
  {Science}\ }\textbf {\bibinfo {volume} {323}},\ \bibinfo {pages} {1184--1187}
  (\bibinfo {year} {2009})}\BibitemShut {NoStop}%
\bibitem [{\citenamefont {Coh}\ \emph {et~al.}(2011)\citenamefont {Coh},
  \citenamefont {Vanderbilt}, \citenamefont {Malashevich},\ and\ \citenamefont
  {I.}}]{Coh_et_al:2011}%
  \BibitemOpen
  \bibfield  {author} {\bibinfo {author} {\bibfnamefont {S.}~\bibnamefont
  {Coh}}, \bibinfo {author} {\bibfnamefont {D.}~\bibnamefont {Vanderbilt}},
  \bibinfo {author} {\bibfnamefont {A.}~\bibnamefont {Malashevich}}, \ and\
  \bibinfo {author} {\bibfnamefont {Souza}\ \bibnamefont {I.}},\ }\bibfield
  {title} {\enquote {\bibinfo {title} {{Chern-Simons orbital magnetoelectric
  coupling in generic insulators}},}\ }\href@noop {} {\bibfield  {journal}
  {\bibinfo  {journal} {Phys. Rev. B}\ }\textbf {\bibinfo {volume} {83}},\
  \bibinfo {pages} {085108} (\bibinfo {year} {2011})}\BibitemShut {NoStop}%
\bibitem [{\citenamefont {Salman}\ \emph {et~al.}()\citenamefont {Salman},
  \citenamefont {Neupert}, \citenamefont {Giblin}, \citenamefont {Kanigel},
  \citenamefont {Morenzoni}, \citenamefont {Prokscha}, \citenamefont
  {Saadaoui}, \citenamefont {Suter},\ and\ \citenamefont
  {Mudry}}]{Salman_et_al:2018}%
  \BibitemOpen
  \bibfield  {author} {\bibinfo {author} {\bibfnamefont {Z.}~\bibnamefont
  {Salman}}, \bibinfo {author} {\bibfnamefont {T.}~\bibnamefont {Neupert}},
  \bibinfo {author} {\bibfnamefont {S.}~\bibnamefont {Giblin}}, \bibinfo
  {author} {\bibfnamefont {A.}~\bibnamefont {Kanigel}}, \bibinfo {author}
  {\bibfnamefont {E.}~\bibnamefont {Morenzoni}}, \bibinfo {author}
  {\bibfnamefont {T.}~\bibnamefont {Prokscha}}, \bibinfo {author}
  {\bibfnamefont {H.}~\bibnamefont {Saadaoui}}, \bibinfo {author}
  {\bibfnamefont {A.}~\bibnamefont {Suter}}, \ and\ \bibinfo {author}
  {\bibfnamefont {C.}~\bibnamefont {Mudry}},\ }\href@noop {} {\enquote
  {\bibinfo {title} {Search for the magnetic monopole image charge in
  topological insulators},}\ }\bibinfo {note} {S$\mu$S Proposal
  20111606}\BibitemShut {NoStop}%
\bibitem [{\citenamefont {Pesin}\ and\ \citenamefont
  {MacDonald}(2013)}]{Pesin/MacDonald:2013}%
  \BibitemOpen
  \bibfield  {author} {\bibinfo {author} {\bibfnamefont {D.~A.}\ \bibnamefont
  {Pesin}}\ and\ \bibinfo {author} {\bibfnamefont {A.~H.}\ \bibnamefont
  {MacDonald}},\ }\bibfield  {title} {\enquote {\bibinfo {title} {Topological
  magnetoelectric effect decay},}\ }\href@noop {} {\bibfield  {journal}
  {\bibinfo  {journal} {Phys. Rev. Lett.}\ }\textbf {\bibinfo {volume} {111}},\
  \bibinfo {pages} {016801} (\bibinfo {year} {2013})}\BibitemShut {NoStop}%
\bibitem [{\citenamefont {Dzyaloshinskii}(1959)}]{dzyalo}%
  \BibitemOpen
  \bibfield  {author} {\bibinfo {author} {\bibfnamefont {I.E.}\ \bibnamefont
  {Dzyaloshinskii}},\ }\href@noop {} {\bibfield  {journal} {\bibinfo  {journal}
  {Zh. Exp. Teor Fiz.}\ }\textbf {\bibinfo {volume} {37}},\ \bibinfo {pages}
  {881} (\bibinfo {year} {1959})}\BibitemShut {NoStop}%
\bibitem [{\citenamefont {Astrov}(1961)}]{astrov}%
  \BibitemOpen
  \bibfield  {author} {\bibinfo {author} {\bibfnamefont {D.N.}\ \bibnamefont
  {Astrov}},\ }\bibfield  {title} {\enquote {\bibinfo {title} {{Magnetoelectric
  effect in chromium oxide}},}\ }\href@noop {} {\bibfield  {journal} {\bibinfo
  {journal} {Sov. Phys. JETP}\ }\textbf {\bibinfo {volume} {13}},\ \bibinfo
  {pages} {729} (\bibinfo {year} {1961})}\BibitemShut {NoStop}%
\bibitem [{\citenamefont {Fiebig}\ \emph {et~al.}(2005)\citenamefont {Fiebig},
  \citenamefont {Pavlov},\ and\ \citenamefont {Pisarev}}]{Fiebig:2005}%
  \BibitemOpen
  \bibfield  {author} {\bibinfo {author} {\bibfnamefont {M.}~\bibnamefont
  {Fiebig}}, \bibinfo {author} {\bibfnamefont {V.V.}\ \bibnamefont {Pavlov}}, \
  and\ \bibinfo {author} {\bibfnamefont {R.V.}\ \bibnamefont {Pisarev}},\
  }\bibfield  {title} {\enquote {\bibinfo {title} {Second-harmonic generation
  as a tool for studying electronic and magnetic structures of crystals:
  review},}\ }\href {\doibase 10.1364/JOSAB.22.000096} {\bibfield  {journal}
  {\bibinfo  {journal} {J. Opt. Soc. Am. B}\ }\textbf {\bibinfo {volume}
  {22}},\ \bibinfo {pages} {96--118} (\bibinfo {year} {2005})}\BibitemShut
  {NoStop}%
\bibitem [{\citenamefont {Coh}\ and\ \citenamefont
  {Vanderbilt}(2014)}]{Coh/Vanderbilt:2014}%
  \BibitemOpen
  \bibfield  {author} {\bibinfo {author} {\bibfnamefont {S.}~\bibnamefont
  {Coh}}\ and\ \bibinfo {author} {\bibfnamefont {D.}~\bibnamefont
  {Vanderbilt}},\ }\bibfield  {title} {\enquote {\bibinfo {title} {Canonical
  magnetic insulators with isotropic magnetoelectric coupling},}\ }\href@noop
  {} {\bibfield  {journal} {\bibinfo  {journal} {Phys. Rev. B}\ }\textbf
  {\bibinfo {volume} {88}},\ \bibinfo {pages} {121106} (\bibinfo {year}
  {2014})}\BibitemShut {NoStop}%
\bibitem [{\citenamefont {Foner}(1963)}]{Foner:1963vi}%
  \BibitemOpen
  \bibfield  {author} {\bibinfo {author} {\bibfnamefont {S.}~\bibnamefont
  {Foner}},\ }\bibfield  {title} {\enquote {\bibinfo {title} {{High-Field
  Antiferromagnetic Resonance in {Cr}$_2$O$_3$}},}\ }\href@noop {} {\bibfield
  {journal} {\bibinfo  {journal} {Phys. Rev.}\ }\textbf {\bibinfo {volume}
  {130}},\ \bibinfo {pages} {183} (\bibinfo {year} {1963})}\BibitemShut
  {NoStop}%
\bibitem [{\citenamefont {Wiegelmann}\ \emph {et~al.}(1994)\citenamefont
  {Wiegelmann}, \citenamefont {Jansen}, \citenamefont {Wyder}, \citenamefont
  {Rivera},\ and\ \citenamefont {Schmid}}]{Wiegelmann:1994tq}%
  \BibitemOpen
  \bibfield  {author} {\bibinfo {author} {\bibfnamefont {H.}~\bibnamefont
  {Wiegelmann}}, \bibinfo {author} {\bibfnamefont {A.G.M.}\ \bibnamefont
  {Jansen}}, \bibinfo {author} {\bibfnamefont {P.}~\bibnamefont {Wyder}},
  \bibinfo {author} {\bibfnamefont {J.~P.}\ \bibnamefont {Rivera}}, \ and\
  \bibinfo {author} {\bibfnamefont {H.}~\bibnamefont {Schmid}},\ }\bibfield
  {title} {\enquote {\bibinfo {title} {{Magnetoelectric effect of Cr2O3 in
  strong static magnetic fields}},}\ }\href@noop {} {\bibfield  {journal}
  {\bibinfo  {journal} {Ferroelectrics}\ }\textbf {\bibinfo {volume} {162}},\
  \bibinfo {pages} {141--146} (\bibinfo {year} {1994})}\BibitemShut {NoStop}%
\bibitem [{\citenamefont {Lal}\ \emph {et~al.}(1967)\citenamefont {Lal},
  \citenamefont {Srivasta},\ and\ \citenamefont {Srivastava}}]{LAL:1967fd}%
  \BibitemOpen
  \bibfield  {author} {\bibinfo {author} {\bibfnamefont {H~B}\ \bibnamefont
  {Lal}}, \bibinfo {author} {\bibfnamefont {R}~\bibnamefont {Srivasta}}, \ and\
  \bibinfo {author} {\bibfnamefont {K.~G.}\ \bibnamefont {Srivastava}},\
  }\bibfield  {title} {\enquote {\bibinfo {title} {{Magnetoelectric Effect in
  Cr2o3 Single Crystal as Studied by Dielectric-Constant Method}},}\
  }\href@noop {} {\bibfield  {journal} {\bibinfo  {journal} {Phys. Rev.}\
  }\textbf {\bibinfo {volume} {154}},\ \bibinfo {pages} {505--{\&}} (\bibinfo
  {year} {1967})}\BibitemShut {NoStop}%
\bibitem [{Note1()}]{Note1}%
  \BibitemOpen
  \bibinfo {note} {While only 11 of the 58 magnetic point groups that allow the
  magnetoelectric effect have uniaxial symmetry, in many of the other cases,
  the tensor can be transformed into a form with a diagonal component and our
  analysis remains relevant for an appropriate choice of surface
  cut.}\BibitemShut {Stop}%
\bibitem [{\citenamefont {Mostovoy}\ \emph {et~al.}(2010)\citenamefont
  {Mostovoy}, \citenamefont {Scaramucci}, \citenamefont {Spaldin},\ and\
  \citenamefont {Delaney}}]{Mostovoy:2010ia}%
  \BibitemOpen
  \bibfield  {author} {\bibinfo {author} {\bibfnamefont {M.}~\bibnamefont
  {Mostovoy}}, \bibinfo {author} {\bibfnamefont {A.}~\bibnamefont
  {Scaramucci}}, \bibinfo {author} {\bibfnamefont {N.~A.}\ \bibnamefont
  {Spaldin}}, \ and\ \bibinfo {author} {\bibfnamefont {K.~T.}\ \bibnamefont
  {Delaney}},\ }\bibfield  {title} {\enquote {\bibinfo {title}
  {{Temperature-Dependent Magnetoelectric Effect from First Principles}},}\
  }\href@noop {} {\bibfield  {journal} {\bibinfo  {journal} {Phys. Rev. Lett.}\
  }\textbf {\bibinfo {volume} {105}},\ \bibinfo {pages} {628} (\bibinfo {year}
  {2010})}\BibitemShut {NoStop}%
\bibitem [{\citenamefont {Morenzoni}\ \emph {et~al.}(1994)\citenamefont
  {Morenzoni}, \citenamefont {Kottmann}, \citenamefont {Maden}, \citenamefont
  {Matthias}, \citenamefont {Meyberg}, \citenamefont {Prokscha}, \citenamefont
  {Wutzke},\ and\ \citenamefont {Zimmermann}}]{Morenzoni1994prl}%
  \BibitemOpen
  \bibfield  {author} {\bibinfo {author} {\bibfnamefont {E.}~\bibnamefont
  {Morenzoni}}, \bibinfo {author} {\bibfnamefont {F.}~\bibnamefont {Kottmann}},
  \bibinfo {author} {\bibfnamefont {D.}~\bibnamefont {Maden}}, \bibinfo
  {author} {\bibfnamefont {B.}~\bibnamefont {Matthias}}, \bibinfo {author}
  {\bibfnamefont {M.}~\bibnamefont {Meyberg}}, \bibinfo {author} {\bibfnamefont
  {Th.}\ \bibnamefont {Prokscha}}, \bibinfo {author} {\bibfnamefont {Th.}\
  \bibnamefont {Wutzke}}, \ and\ \bibinfo {author} {\bibfnamefont
  {U.}~\bibnamefont {Zimmermann}},\ }\bibfield  {title} {\enquote {\bibinfo
  {title} {Generation of very slow polarized positive muons},}\ }\href
  {\doibase 10.1103/PhysRevLett.72.2793} {\bibfield  {journal} {\bibinfo
  {journal} {Phys. Rev. Lett.}\ }\textbf {\bibinfo {volume} {72}},\ \bibinfo
  {pages} {2793} (\bibinfo {year} {1994})}\BibitemShut {NoStop}%
\bibitem [{\citenamefont {Prokscha}\ \emph {et~al.}(2008)\citenamefont
  {Prokscha}, \citenamefont {Morenzoni}, \citenamefont {Deiters}, \citenamefont
  {Foroughi}, \citenamefont {George}, \citenamefont {Kobler}, \citenamefont
  {Suter},\ and\ \citenamefont {Vrankovic}}]{Prokscha2008nima}%
  \BibitemOpen
  \bibfield  {author} {\bibinfo {author} {\bibfnamefont {T.}~\bibnamefont
  {Prokscha}}, \bibinfo {author} {\bibfnamefont {E.}~\bibnamefont {Morenzoni}},
  \bibinfo {author} {\bibfnamefont {K.}~\bibnamefont {Deiters}}, \bibinfo
  {author} {\bibfnamefont {F.}~\bibnamefont {Foroughi}}, \bibinfo {author}
  {\bibfnamefont {D.}~\bibnamefont {George}}, \bibinfo {author} {\bibfnamefont
  {R.}~\bibnamefont {Kobler}}, \bibinfo {author} {\bibfnamefont
  {A.}~\bibnamefont {Suter}}, \ and\ \bibinfo {author} {\bibfnamefont
  {V.}~\bibnamefont {Vrankovic}},\ }\bibfield  {title} {\enquote {\bibinfo
  {title} {The new μe4 beam at {PSI}: A hybrid-type large acceptance channel
  for the generation of a high intensity surface-muon beam},}\ }\href {\doibase
  10.1016/j.nima.2008.07.081} {\bibfield  {journal} {\bibinfo  {journal} {Nuc.
  Inst. Phys. A}\ }\textbf {\bibinfo {volume} {595}},\ \bibinfo {pages}
  {317--331} (\bibinfo {year} {2008})}\BibitemShut {NoStop}%
\bibitem [{\citenamefont {Morenzoni}\ \emph {et~al.}(2000)\citenamefont
  {Morenzoni}, \citenamefont {Gl\"uckler}, \citenamefont {Prokscha},
  \citenamefont {Weber}, \citenamefont {Forgan}, \citenamefont {Jackson},
  \citenamefont {Luetkens}, \citenamefont {Niedermayer}, \citenamefont
  {Pleines}, \citenamefont {Birke}, \citenamefont {Hofer}, \citenamefont
  {Litterst}, \citenamefont {Riseman},\ and\ \citenamefont
  {Schatz}}]{morenzoni-2000}%
  \BibitemOpen
  \bibfield  {author} {\bibinfo {author} {\bibfnamefont {E.}~\bibnamefont
  {Morenzoni}}, \bibinfo {author} {\bibfnamefont {H.}~\bibnamefont
  {Gl\"uckler}}, \bibinfo {author} {\bibfnamefont {T.}~\bibnamefont
  {Prokscha}}, \bibinfo {author} {\bibfnamefont {H.P.}\ \bibnamefont {Weber}},
  \bibinfo {author} {\bibfnamefont {E.M.}\ \bibnamefont {Forgan}}, \bibinfo
  {author} {\bibfnamefont {T.J.}\ \bibnamefont {Jackson}}, \bibinfo {author}
  {\bibfnamefont {H.}~\bibnamefont {Luetkens}}, \bibinfo {author}
  {\bibfnamefont {C.}~\bibnamefont {Niedermayer}}, \bibinfo {author}
  {\bibfnamefont {M.}~\bibnamefont {Pleines}}, \bibinfo {author} {\bibfnamefont
  {M.}~\bibnamefont {Birke}}, \bibinfo {author} {\bibfnamefont
  {A.}~\bibnamefont {Hofer}}, \bibinfo {author} {\bibfnamefont
  {J.}~\bibnamefont {Litterst}}, \bibinfo {author} {\bibfnamefont
  {T.}~\bibnamefont {Riseman}}, \ and\ \bibinfo {author} {\bibfnamefont
  {G.}~\bibnamefont {Schatz}},\ }\bibfield  {title} {\enquote {\bibinfo {title}
  {Low-energy μsr at psi: present and future},}\ }\href {\doibase
  https://doi.org/10.1016/S0921-4526(00)00303-3} {\bibfield  {journal}
  {\bibinfo  {journal} {Physica B: Condens. Matter}\ }\textbf {\bibinfo
  {volume} {289-290}},\ \bibinfo {pages} {653 -- 657} (\bibinfo {year}
  {2000})}\BibitemShut {NoStop}%
\bibitem [{\citenamefont {Morenzoni}\ \emph {et~al.}(2004)\citenamefont
  {Morenzoni}, \citenamefont {Prokscha}, \citenamefont {Suter}, \citenamefont
  {Luetkens},\ and\ \citenamefont {Khasanov}}]{morenzoni-2004}%
  \BibitemOpen
  \bibfield  {author} {\bibinfo {author} {\bibfnamefont {E.}~\bibnamefont
  {Morenzoni}}, \bibinfo {author} {\bibfnamefont {T.}~\bibnamefont {Prokscha}},
  \bibinfo {author} {\bibfnamefont {A.}~\bibnamefont {Suter}}, \bibinfo
  {author} {\bibfnamefont {H.}~\bibnamefont {Luetkens}}, \ and\ \bibinfo
  {author} {\bibfnamefont {R.}~\bibnamefont {Khasanov}},\ }\bibfield  {title}
  {\enquote {\bibinfo {title} {Nano-scale thin film investigations with slow
  polarized muons},}\ }\href {http://stacks.iop.org/0953-8984/16/i=40/a=010}
  {\bibfield  {journal} {\bibinfo  {journal} {J. Phys. Condens. Matter}\
  }\textbf {\bibinfo {volume} {16}},\ \bibinfo {pages} {S4583} (\bibinfo {year}
  {2004})}\BibitemShut {NoStop}%
\bibitem [{\citenamefont {Yaouanc}\ and\ \citenamefont
  {de~R\'eotier}(2010)}]{Yaouanc2010}%
  \BibitemOpen
  \bibfield  {author} {\bibinfo {author} {\bibfnamefont {A.n}\ \bibnamefont
  {Yaouanc}}\ and\ \bibinfo {author} {\bibfnamefont {P.~D.}\ \bibnamefont
  {de~R\'eotier}},\ }\href@noop {} {\emph {\bibinfo {title} {Muon Spin
  Rotation, Relaxation, and Resonance: Applications to Condensed Matter}}}\
  (\bibinfo  {publisher} {{OUP} Oxford},\ \bibinfo {year} {2010})\BibitemShut
  {NoStop}%
\bibitem [{\citenamefont {Al-Mahdawi}\ \emph {et~al.}(2017)\citenamefont
  {Al-Mahdawi}, \citenamefont {Pati}, \citenamefont {Shiokawa}, \citenamefont
  {Ye}, \citenamefont {Nozaki},\ and\ \citenamefont {Sahashi}}]{mahdawi-2017}%
  \BibitemOpen
  \bibfield  {author} {\bibinfo {author} {\bibfnamefont {M.}~\bibnamefont
  {Al-Mahdawi}}, \bibinfo {author} {\bibfnamefont {S.~P.}\ \bibnamefont
  {Pati}}, \bibinfo {author} {\bibfnamefont {Y.}~\bibnamefont {Shiokawa}},
  \bibinfo {author} {\bibfnamefont {S.}~\bibnamefont {Ye}}, \bibinfo {author}
  {\bibfnamefont {T.}~\bibnamefont {Nozaki}}, \ and\ \bibinfo {author}
  {\bibfnamefont {M.}~\bibnamefont {Sahashi}},\ }\bibfield  {title} {\enquote
  {\bibinfo {title} {Low-energy magnetoelectric control of domain states in
  exchange-coupled heterostructures},}\ }\href {\doibase
  10.1103/PhysRevB.95.144423} {\bibfield  {journal} {\bibinfo  {journal} {Phys.
  Rev. B}\ }\textbf {\bibinfo {volume} {95}},\ \bibinfo {pages} {144423}
  (\bibinfo {year} {2017})}\BibitemShut {NoStop}%
\bibitem [{\citenamefont {Prokscha}\ \emph {et~al.}(2007)\citenamefont
  {Prokscha}, \citenamefont {Morenzoni}, \citenamefont {Eshchenko},
  \citenamefont {Garifianov}, \citenamefont {Gl\"uckler}, \citenamefont
  {Khasanov}, \citenamefont {Luetkens},\ and\ \citenamefont
  {Suter}}]{PhysRevLett.98.227401}%
  \BibitemOpen
  \bibfield  {author} {\bibinfo {author} {\bibfnamefont {T.}~\bibnamefont
  {Prokscha}}, \bibinfo {author} {\bibfnamefont {E.}~\bibnamefont {Morenzoni}},
  \bibinfo {author} {\bibfnamefont {D.~G.}\ \bibnamefont {Eshchenko}}, \bibinfo
  {author} {\bibfnamefont {N.}~\bibnamefont {Garifianov}}, \bibinfo {author}
  {\bibfnamefont {H.}~\bibnamefont {Gl\"uckler}}, \bibinfo {author}
  {\bibfnamefont {R.}~\bibnamefont {Khasanov}}, \bibinfo {author}
  {\bibfnamefont {H.}~\bibnamefont {Luetkens}}, \ and\ \bibinfo {author}
  {\bibfnamefont {A.}~\bibnamefont {Suter}},\ }\bibfield  {title} {\enquote
  {\bibinfo {title} {Formation of hydrogen impurity states in silicon and
  insulators at low implantation energies},}\ }\href {\doibase
  10.1103/PhysRevLett.98.227401} {\bibfield  {journal} {\bibinfo  {journal}
  {Phys. Rev. Lett.}\ }\textbf {\bibinfo {volume} {98}},\ \bibinfo {pages}
  {227401} (\bibinfo {year} {2007})}\BibitemShut {NoStop}%
\bibitem [{\citenamefont {Borisov}\ \emph {et~al.}(2016)\citenamefont
  {Borisov}, \citenamefont {Ashida}, \citenamefont {Nozaki}, \citenamefont
  {Sahashi},\ and\ \citenamefont {Lederman}}]{Borisov2016}%
  \BibitemOpen
  \bibfield  {author} {\bibinfo {author} {\bibfnamefont {P.}~\bibnamefont
  {Borisov}}, \bibinfo {author} {\bibfnamefont {T.}~\bibnamefont {Ashida}},
  \bibinfo {author} {\bibfnamefont {T.}~\bibnamefont {Nozaki}}, \bibinfo
  {author} {\bibfnamefont {M.}~\bibnamefont {Sahashi}}, \ and\ \bibinfo
  {author} {\bibfnamefont {D.}~\bibnamefont {Lederman}},\ }\bibfield  {title}
  {\enquote {\bibinfo {title} {Magnetoelectric properties of 500-nm
  {Cr}$_2${O}$_3$ films},}\ }\href {\doibase 10.1103/PhysRevB.93.174415}
  {\bibfield  {journal} {\bibinfo  {journal} {Phys. Rev. B}\ }\textbf {\bibinfo
  {volume} {93}},\ \bibinfo {pages} {174415} (\bibinfo {year}
  {2016})}\BibitemShut {NoStop}%
\bibitem [{\citenamefont {Morenzoni}\ \emph {et~al.}(2002)\citenamefont
  {Morenzoni}, \citenamefont {{Gl\"uckler}}, \citenamefont {Prokscha},
  \citenamefont {Khasanov}, \citenamefont {Luetkens}, \citenamefont {Birke},
  \citenamefont {Forgan}, \citenamefont {Niedermayer},\ and\ \citenamefont
  {Pleines}}]{Morenzoni2002nimb}%
  \BibitemOpen
  \bibfield  {author} {\bibinfo {author} {\bibfnamefont {E.}~\bibnamefont
  {Morenzoni}}, \bibinfo {author} {\bibfnamefont {H.}~\bibnamefont
  {{Gl\"uckler}}}, \bibinfo {author} {\bibfnamefont {T.}~\bibnamefont
  {Prokscha}}, \bibinfo {author} {\bibfnamefont {R.}~\bibnamefont {Khasanov}},
  \bibinfo {author} {\bibfnamefont {H.}~\bibnamefont {Luetkens}}, \bibinfo
  {author} {\bibfnamefont {M.}~\bibnamefont {Birke}}, \bibinfo {author}
  {\bibfnamefont {E.~M.}\ \bibnamefont {Forgan}}, \bibinfo {author}
  {\bibfnamefont {Ch.}\ \bibnamefont {Niedermayer}}, \ and\ \bibinfo {author}
  {\bibfnamefont {M.}~\bibnamefont {Pleines}},\ }\bibfield  {title} {\enquote
  {\bibinfo {title} {Implantation studies of {keV} positive muons in thin
  metallic layers},}\ }\href {\doibase 10.1016/S0168-583X(01)01166-1}
  {\bibfield  {journal} {\bibinfo  {journal} {Nucl. Instr. Meth. Phys. Res. B}\
  }\textbf {\bibinfo {volume} {192}},\ \bibinfo {pages} {254--266} (\bibinfo
  {year} {2002})}\BibitemShut {NoStop}%
\bibitem [{\citenamefont {Appel}\ \emph {et~al.}()\citenamefont {Appel},
  \citenamefont {Shields}, \citenamefont {Kosub}, \citenamefont {H\"ubner},
  \citenamefont {Fassbender}, \citenamefont {Makarov},\ and\ \citenamefont
  {Maletinsky}}]{Appel_et_al:2018}%
  \BibitemOpen
  \bibfield  {author} {\bibinfo {author} {\bibfnamefont {P.}~\bibnamefont
  {Appel}}, \bibinfo {author} {\bibfnamefont {B.~J.}\ \bibnamefont {Shields}},
  \bibinfo {author} {\bibfnamefont {T.}~\bibnamefont {Kosub}}, \bibinfo
  {author} {\bibfnamefont {R.}~\bibnamefont {H\"ubner}}, \bibinfo {author}
  {\bibfnamefont {J.}~\bibnamefont {Fassbender}}, \bibinfo {author}
  {\bibfnamefont {D.}~\bibnamefont {Makarov}}, \ and\ \bibinfo {author}
  {\bibfnamefont {P.}~\bibnamefont {Maletinsky}},\ }\href@noop {} {\enquote
  {\bibinfo {title} {Nanomagnetism of magnetoelectric granular thin-film
  antiferromagnets},}\ }\bibinfo {note} {{arXiv}:1806.02572 (2018)}\BibitemShut
  {NoStop}%
\bibitem [{\citenamefont {Andreev}(1996)}]{Andreev1996}%
  \BibitemOpen
  \bibfield  {author} {\bibinfo {author} {\bibfnamefont {A.~F.}\ \bibnamefont
  {Andreev}},\ }\bibfield  {title} {\enquote {\bibinfo {title} {Macroscopic
  magnetic fields of antiferromagnets},}\ }\href {\doibase 10.1134/1.566978}
  {\bibfield  {journal} {\bibinfo  {journal} {J. Exp. Theor. Phys. Lett}\
  }\textbf {\bibinfo {volume} {63}},\ \bibinfo {pages} {758--762} (\bibinfo
  {year} {1996})}\BibitemShut {NoStop}%
\bibitem [{\citenamefont {Astrov}\ \emph {et~al.}(1996)\citenamefont {Astrov},
  \citenamefont {Ermakov}, \citenamefont {Borovik-Romanov}, \citenamefont
  {Kolevatov},\ and\ \citenamefont {Nizhankovskii}}]{Astrov1996}%
  \BibitemOpen
  \bibfield  {author} {\bibinfo {author} {\bibfnamefont {D.~N.}\ \bibnamefont
  {Astrov}}, \bibinfo {author} {\bibfnamefont {N.~B.}\ \bibnamefont {Ermakov}},
  \bibinfo {author} {\bibfnamefont {A.~S.}\ \bibnamefont {Borovik-Romanov}},
  \bibinfo {author} {\bibfnamefont {E.~G.}\ \bibnamefont {Kolevatov}}, \ and\
  \bibinfo {author} {\bibfnamefont {V.~I.}\ \bibnamefont {Nizhankovskii}},\
  }\bibfield  {title} {\enquote {\bibinfo {title} {External quadrupole magnetic
  field of antiferromagnetic {Cr}$_2${O}$_3$},}\ }\href {\doibase
  10.1134/1.566976} {\bibfield  {journal} {\bibinfo  {journal} {J. Exp. Theor.
  Phys. Lett}\ }\textbf {\bibinfo {volume} {63}},\ \bibinfo {pages} {745--751}
  (\bibinfo {year} {1996})}\BibitemShut {NoStop}%
\bibitem [{\citenamefont {Fallarino}\ \emph {et~al.}(2015)\citenamefont
  {Fallarino}, \citenamefont {Berger},\ and\ \citenamefont
  {Binek}}]{fallarino}%
  \BibitemOpen
  \bibfield  {author} {\bibinfo {author} {\bibfnamefont {L.}~\bibnamefont
  {Fallarino}}, \bibinfo {author} {\bibfnamefont {A.}~\bibnamefont {Berger}}, \
  and\ \bibinfo {author} {\bibfnamefont {C.}~\bibnamefont {Binek}},\ }\bibfield
   {title} {\enquote {\bibinfo {title} {Magnetic field induced switching of the
  antiferromagnetic order parameter in thin films of magnetoelectric
  chromia},}\ }\href {\doibase 10.1103/PhysRevB.91.054414} {\bibfield
  {journal} {\bibinfo  {journal} {Phys. Rev. B}\ }\textbf {\bibinfo {volume}
  {91}},\ \bibinfo {pages} {054414} (\bibinfo {year} {2015})}\BibitemShut
  {NoStop}%
\bibitem [{\citenamefont {Schoenherr}\ \emph {et~al.}(2017)\citenamefont
  {Schoenherr}, \citenamefont {Giraldo}, \citenamefont {Lilienblum},
  \citenamefont {Trassin}, \citenamefont {Meier},\ and\ \citenamefont
  {Fiebig}}]{Schoenherr_et_al:2017}%
  \BibitemOpen
  \bibfield  {author} {\bibinfo {author} {\bibfnamefont {P.}~\bibnamefont
  {Schoenherr}}, \bibinfo {author} {\bibfnamefont {L.}~\bibnamefont {Giraldo},
  \bibfnamefont {M.}}, \bibinfo {author} {\bibfnamefont {M.}~\bibnamefont
  {Lilienblum}}, \bibinfo {author} {\bibfnamefont {M.}~\bibnamefont {Trassin}},
  \bibinfo {author} {\bibfnamefont {D.}~\bibnamefont {Meier}}, \ and\ \bibinfo
  {author} {\bibfnamefont {M.}~\bibnamefont {Fiebig}},\ }\bibfield  {title}
  {\enquote {\bibinfo {title} {Magnetoelectric force microscopy on
  antiferromagnetic 180$^{\circ}$ domains in {Cr$_2$O$_3$}},}\ }\href@noop {}
  {\bibfield  {journal} {\bibinfo  {journal} {Materials}\ }\textbf {\bibinfo
  {volume} {10}},\ \bibinfo {pages} {1051} (\bibinfo {year}
  {2017})}\BibitemShut {NoStop}%
\end{thebibliography}%

\newpage
\appendix

\section{Detailed solution for the isotropic case}\label{app:iso}

In this and following appendices we use cgs units for conciseness of notation.
From the electrostatic boundary conditions and equations \eqref{eq::pot1} and \eqref{eq::pot2} it follows that the image monopole, $m''=m'$ and that the image charge, $q''=q+q'$
\begin{eqnarray*}
m''&=&m'\\
q''&=&q+q'\quad .
\end{eqnarray*}

From the second and third boundary condition it is found that
\begin{eqnarray*}
\epsilon q'' + \alpha m &=& q - q'\\
\mu m' + \alpha q'' &=& -m \quad .
\end{eqnarray*}

From this we find the following equations
\begin{equation*}
\begin{pmatrix} 1 & -1 & 0 \\ 1 &\epsilon & \alpha \\ 0 &\alpha & \mu+1\end{pmatrix}\myvec{q'}{q''}{m'}=\myvec{-q}{q}{0}\quad ,
\end{equation*}
which we solve using Gaussian transformations:

\begin{equation*}
\begin{pmatrix} 1 & -1 & 0 & -q\\  &\epsilon+1 & \alpha &2q \\ 0 & 0 & \mu+1 - \dfrac{\alpha^2}{\epsilon+1}& -\dfrac{2q\alpha}{\epsilon+1}\end{pmatrix}\quad .
\end{equation*}
We find
\begin{equation*}
m'=-\dfrac{2q\alpha}{(\mu+1)(\epsilon+1)-\alpha^2} \quad ,
\end{equation*}
\begin{equation*}
q''=\dfrac{1}{\epsilon+1}\left(+\dfrac{2q\alpha^2}{(\mu+1)(\epsilon+1)-\alpha^2} +2q\right)=\dfrac{2q(\mu+1)}{(\mu+1)(\epsilon+1)-\alpha^2} \quad ,
\end{equation*}
and 
\begin{equation*}
q'=-q+\dfrac{2q(\mu+1)}{(\mu+1)(\epsilon+1)-\alpha^2}=-\dfrac{q(\mu+1)(\epsilon-1)-\alpha^2}{(\mu+1)(\epsilon+1)-\alpha^2} \quad .
\end{equation*}

Using the previous results one finds the potentials
\begin{eqnarray*}
\phi_e^{out} &=& \dfrac{q}{\left|\vec{r}-\vec{r_1}\right|}-\dfrac{q}{\abs{\vec{r}-\vec{r_2}}}\dfrac{(\mu+1)(\epsilon-1)-\alpha^2}{(\mu+1)(\epsilon+1)-\alpha^2} \\
\phi^{out}_m &=& -\dfrac{q}{\abs{\vec r - \vec r_2}}\dfrac{2\alpha}{(\mu+1)(\epsilon+1)-\alpha^2}\\
\phi_e^{in} &=& \dfrac{q}{\abs{\vec r- \vec r_1}}\dfrac{2(\mu+1)}{(\mu+1)(\epsilon+1)-\alpha^2}\\
\phi^{in}_m &=& -\dfrac{q}{\abs{\vec r - \vec r_1}}\dfrac{2\alpha}{(\mu+1)(\epsilon+1)-\alpha^2} \quad ,
\end{eqnarray*}

where $\phi$ are electric $(e)$ and magnetic $(m)$ potentials inside $(i)$ and outside $(o)$ the magnetoelectric slab. Taking the gradients, this leads to the fields:

\begin{eqnarray*}
\vec E^{out}(\vec r)&=&\dfrac{q (\vec r - \vec r_1)}{\abs{\vec r-\vec r_1}^{3/2}}-\dfrac{q (\vec r - \vec r_2)}{\abs{\vec{r}-\vec{r_2}}^{3/2}}\dfrac{(\mu+1)(\epsilon-1)-\alpha^2}{(\mu+1)(\epsilon+1)-\alpha^2}\\
\vec H^{out}(\vec r) &=& -\dfrac{q(\vec r - \vec r_2)}{\abs{\vec r - \vec r_2}^{3/2}}\dfrac{2\alpha}{(\mu+1)(\epsilon+1)-\alpha^2}\\
\vec E^{in}(\vec r)&=&\dfrac{q(\vec r - \vec r_1)}{\abs{\vec r- \vec r_1}^{3/2}}\dfrac{2(\mu+1)}{(\mu+1)(\epsilon+1)-\alpha^2}\\
\vec H^{in}(\vec r) &=& -\dfrac{q(\vec r - \vec r_1)}{\abs{\vec r - \vec r_1}^{3/2}}\dfrac{2\alpha}{(\mu+1)(\epsilon+1)-\alpha^2} \quad .
\end{eqnarray*}

\section{Detailed solution for the uniaxial case}\label{app:aniso}
To solve the problem of a charge adjacent to  a slab of uniaxial material one starts with the coupled equations inside the magnetoelectric in the absence of free charge. Again we use cgs units for conciseness.
\begin{align*}
(\epsilon_\perp\nabla_\perp + \epsilon_\parallel\nabla_\parallel)\vc E+(\alpha_\perp\nabla_\perp+
\alpha_\parallel\nabla_\parallel)\vc H&=0\\
(\mu_\perp \nabla_\perp+\mu_\parallel\nabla_\perp) \vc H+(\alpha_\perp \nabla_\perp+\alpha_\parallel\nabla_\parallel)\vc E &=0
\end{align*}
where $\nabla_\parallel = \begin{pmatrix}0 \\ 0 \\\frac{\partial}{\partial z}\end{pmatrix}$ and $\nabla_\perp= \begin{pmatrix}\frac{\partial}{\partial x}\\ \frac{\partial}{\partial y} \\ 0\end{pmatrix}$ are parts of the $\nabla$-operator which are anti-parallel and parallel to the anisotropy axis.\\

Taking the partial fourier transform along x and y, defined by
\begin{align*}
F(x,y,z)&=\dfrac{1}{4\pi^2}\int\int dk_xdk_y F(k_x,k_y,z)e^{ik_xx}e^{ik_yy}\\
F(k_x,k_y,z)&= \int \int dx dy F(x,y,z) e^{-ik_xx}e^{-ik_yy}\quad ,
\end{align*}
we obtain the Fourier transformed magnetoelectric differential equations in terms of magnetic and electric potentials 
\begin{align*}
\begin{pmatrix}
\mu_\parallel & \alpha_\parallel \\
\alpha_\parallel & \epsilon_\parallel
\end{pmatrix}\begin{pmatrix}\phi_m'' \\ \phi_e''\end{pmatrix}=k^2\begin{pmatrix}\mu_\perp & \alpha_\perp\\
\alpha_\perp & \epsilon_\perp
\end{pmatrix}
\begin{pmatrix}\phi_m \\ \phi_e\end{pmatrix}\quad,
\end{align*}
where $k^2=k_x^2+k_y^2$ and the $'$ indicates the derivative with respect to $z$. Multiplying with the inverse of the first matrix and diagonalizing the equation we find that

\begin{align*}
\begin{pmatrix}\phi_m'' \\ \phi_e''\end{pmatrix}=\dfrac{k^2}{\mu_\parallel \epsilon_\parallel-\alpha_\parallel^2}
\begin{pmatrix} \mu_\perp\epsilon_\parallel -\alpha_\parallel \alpha_\perp & \epsilon_\parallel \alpha_\perp-\alpha_\parallel \epsilon_\perp\\
-\alpha_\parallel\mu_\perp+\mu_\parallel
\alpha_\perp & \epsilon_\perp \mu_\parallel -\alpha_\parallel\alpha_\perp
\end{pmatrix}\begin{pmatrix}\phi_m\\ \phi_e\end{pmatrix}.
\end{align*}

Diagonalizing this equation we obtain the eigenvalues
\begin{align*}
\lambda_1 &= -k\frac{\sqrt{-\gamma+a+d}}{\sqrt{2}}\\
\lambda_2 &= k\frac{\sqrt{-\gamma+a+d}}{\sqrt{2}}\\
\lambda_3 &= -k\frac{\sqrt{\gamma+a+d}}{\sqrt{2}}\\
\lambda_4 &= k\frac{\sqrt{\gamma+a+d}}{\sqrt{2}}\quad ,
\end{align*}
where we substituted

\begin{align*}
a&=\dfrac{\mu_\perp\epsilon_\parallel -\alpha_\parallel \alpha_\perp}{\mu_\parallel \epsilon_\parallel-\alpha_\parallel^2}\\
b&=\dfrac{\epsilon_\parallel \alpha_\perp-\alpha_\parallel \epsilon_\perp}{\mu_\parallel \epsilon_\parallel-\alpha_\parallel^2}\\
c&=\dfrac{-\alpha_\parallel\mu_\perp+\mu_\parallel
\alpha_\perp}{\mu_\parallel \epsilon_\parallel-\alpha_\parallel^2}\\
d&=\dfrac{\epsilon_\perp \mu_\parallel -\alpha_\parallel\alpha_\perp}{\mu_\parallel \epsilon_\parallel-\alpha_\parallel^2}\\
\gamma&=\sqrt{a^2-2 a d+4 b c+d^2} \quad .
\end{align*}
The eigenvectors are given by:
\begin{align*}
\vc v_1&=\begin{pmatrix}
\frac{1}{k}\frac{\sqrt{2} b \sqrt{a+d-\gamma }}{d (a-d+\gamma )-2 b c}   \\
 -\frac{-a+d+\gamma }{2 c}  \\
 -\frac{1}{k}\frac{\sqrt{2}}{\sqrt{a+d-\gamma }} \\
1 \\
\end{pmatrix}\quad ,
&
\vc v_2&=\begin{pmatrix}
 -\frac{1}{k}\frac{\sqrt{2} b \sqrt{a+d-\gamma }}{d (a-d+\gamma )-2 b c}\\
 \frac{a-d+\gamma }{2 c}\\
  -\frac{1}{k}\frac{\sqrt{2}}{\sqrt{a+d+\gamma }}\\
 1
\end{pmatrix}\quad ,
\\
\vc v_3&=\begin{pmatrix}
-\frac{1}{k}\frac{\sqrt{2} b \sqrt{a+d+\gamma }}{2 b c+d (-a+d+\gamma )}\\
-\frac{-a+d+\gamma }{2 c}\\
\frac{1}{k}\frac{\sqrt{2}}{\sqrt{a+d-\gamma }}\\
 1
\end{pmatrix}\quad ,
&
\vc v_4&=\begin{pmatrix}
\frac{1}{k}\frac{\sqrt{2} b \sqrt{a+d+\gamma }}{2 b c+d (-a+d+\gamma )}\\
\frac{a-d+\gamma }{2 c}\\
\frac{1}{k}\frac{\sqrt{2}}{\sqrt{a+d+\gamma }}\\
1
\end{pmatrix}\quad .
\end{align*}
Since the potential should not diverge for $z\to-\infty$, $C_1$ and $C_3$ are zero, which means that the solution can be written as a combination of the second and the fourth eigenfunctions $\vc v_2$ and $\vc v_4$:\\
\begin{align*}
\phi_m&=-C_2\frac{1}{k}\frac{\sqrt{2} b \sqrt{a+d-\gamma }}{d (a-d+\gamma )-2 b c}e^{k\frac{\sqrt{-\gamma+a+d}}{\sqrt{2}}z}\\&+C_4\frac{1}{k}\frac{\sqrt{2} b \sqrt{a+d+\gamma }}{2 b c+d (-a+d+\gamma )}e^{k\frac{\sqrt{\gamma+a+d}}{\sqrt{2}}z}\\
\phi_e&=C_2\frac{1}{k}\frac{\sqrt{2}}{\sqrt{a+d-\gamma }}e^{k\frac{\sqrt{-\gamma+a+d}}{\sqrt{2}}z}\\&+C_4\frac{1}{k}\frac{\sqrt{2}}{\sqrt{a+d+\gamma }}e^{k\frac{\sqrt{\gamma+a+d}}{\sqrt{2}}z}\quad .
\end{align*}
In the vacuum half space the Maxwell equations reduce to
\begin{align*}
\nabla \cdot \vc E &= 4\pi q \delta(\vc r-\vc r_0)\\
\nabla \cdot \vc H &= 0 \quad .
\end{align*}
Fourier transforming in the $xy$-plane we obtain
\begin{align*}
\nabla^2 \phi_e(k_x,k_y,z)&-(k_x^2+k_y^2)\phi_e(k_x,k_y,z) =  4\pi q \delta(z-z_0)\\
\nabla^2\phi_m(k_x,k_y,z)& -(k_x^2+k_y^2)\phi_m(k_x,k_z,z)= 0
\end{align*}
The general solutions to these equations in fourier space are given by
\begin{align*}
\phi^{vac}_e&=D_1 e^{-k(z+z_0)}+\dfrac{2\pi q}{k}e^{-k|z-z0|}\\
\phi^{vac}_m&=D_2 e^{-k(z+z_0)} \quad .
\end{align*}
Applying the inverse fourier transform we obtain

\begin{align*}
\phi_e^{out}(x,y,z)&=\underbrace{\dfrac{q}{\sqrt{x^2+y^2+|z-z_0|^2}}}_{\text{Potential of the point charge}}+\underbrace{\dfrac{1}{2\pi}\dfrac{D_2}{\sqrt{x^2+y^2+|z+z_0|^2}}}_\text{Electric image charge}\\
\phi_m^{out}(x,y,z)&=\underbrace{\dfrac{1}{2\pi}\dfrac{D_1}{\sqrt{x^2+y^2+|z+z_0|^2}}}_\text{Magnetic image charge}
\end{align*}
outside the material, and inside the material,
\begin{align*}
\phi_m^{in}(x,y,z) &=-\dfrac{C_2}{2\pi}\frac{\sqrt{2} b \sqrt{a+d-\gamma }}{d (a-d+\gamma )-2 b c}\dfrac{1}{\sqrt{x^2+y^2+|\frac{\sqrt{-\gamma+a+d}}{\sqrt{2}}z-z_0|^2}}\nonumber\\
&+\dfrac{C_4}{2\pi}\frac{\sqrt{2} b \sqrt{a+d+\gamma }}{2 b c+d (-a+d+\gamma )}\dfrac{1}{\sqrt{x^2+y^2+|\frac{\sqrt{\gamma+a+d}}{\sqrt{2}}z-z_0|^2}}\\
\phi^{in}_e(x,y,z)&=\dfrac{C_2}{2\pi}\frac{\sqrt{2}}{\sqrt{a+d-\gamma }}\dfrac{1}{\sqrt{x^2+y^2+|\frac{\sqrt{-\gamma+a+d}}{\sqrt{2}}z-z_0|^2}}\nonumber\\&+\dfrac{C_4}{2\pi}\frac{\sqrt{2}}{\sqrt{a+d+\gamma }}\dfrac{1}{\sqrt{x^2+y^2+|\frac{\sqrt{\gamma+a+d}}{\sqrt{2}}z-z_0|^2}}\quad .
\end{align*}

One now can solve the system of equations for the constants by imposing the electromagnetic boundary conditions.

\section{Effect of anisotropy on the monopolar field}

From Eqn.~\eqref{eq::exact} we see that the strength of the magnetic monopolar field is determined by the parameter $c^{out}_{b1}$, which has a functional dependence on the three tensors $\mat{\epsilon},\mat{\mu}$ and $\mat{\alpha}$. To understand this dependence we next analyze the magnitude of  $c^{out}_{b1}$ as we vary the three response functions individually. 

\begin{figure}
\centering
\begin{minipage}{\columnwidth}
\hspace{-150pt} a)\\
\includegraphics[width=0.55\columnwidth]{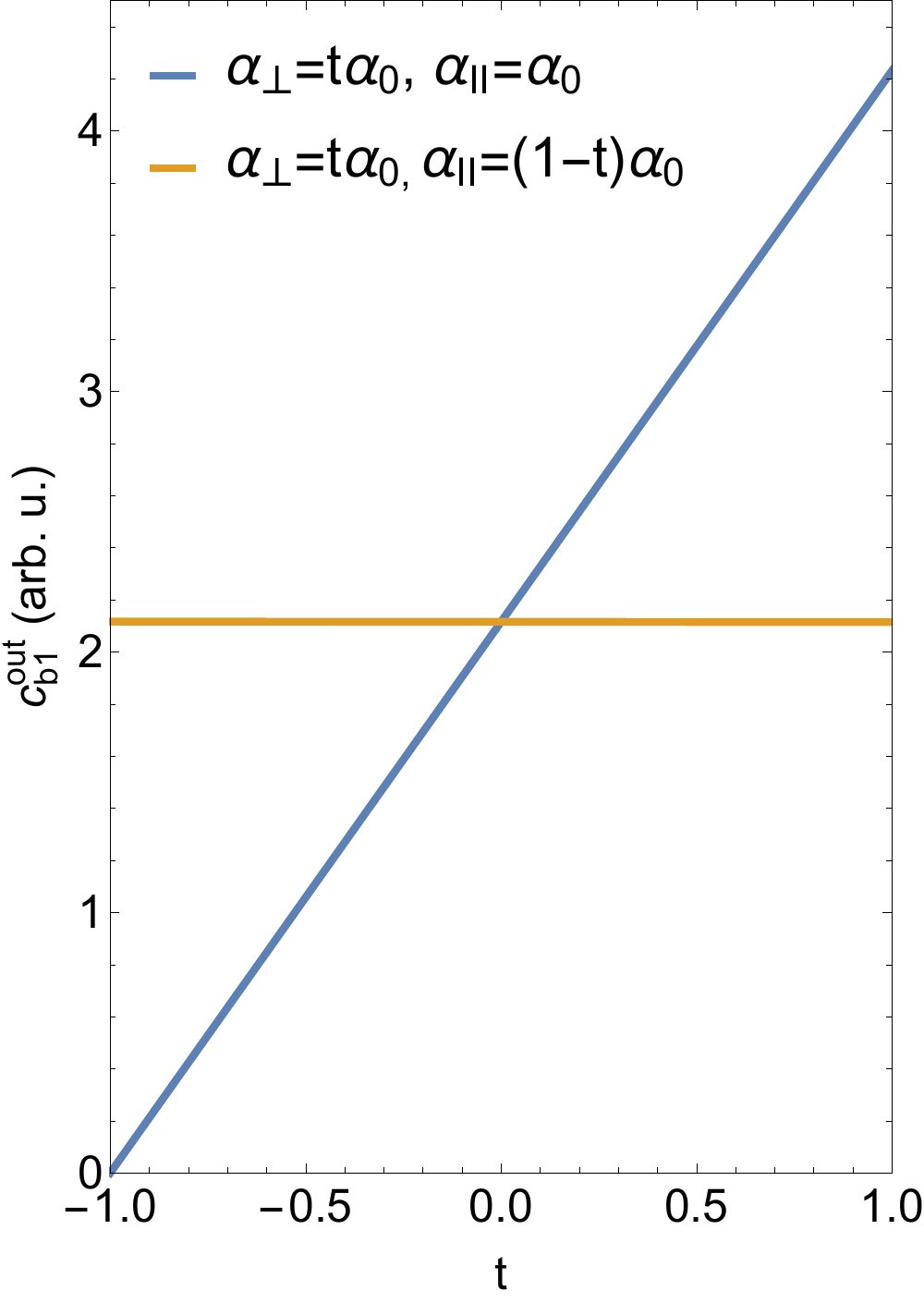}
\end{minipage}
\begin{minipage}{0.9\columnwidth}\hspace{-150pt} b)\\\includegraphics[width=0.6\columnwidth]{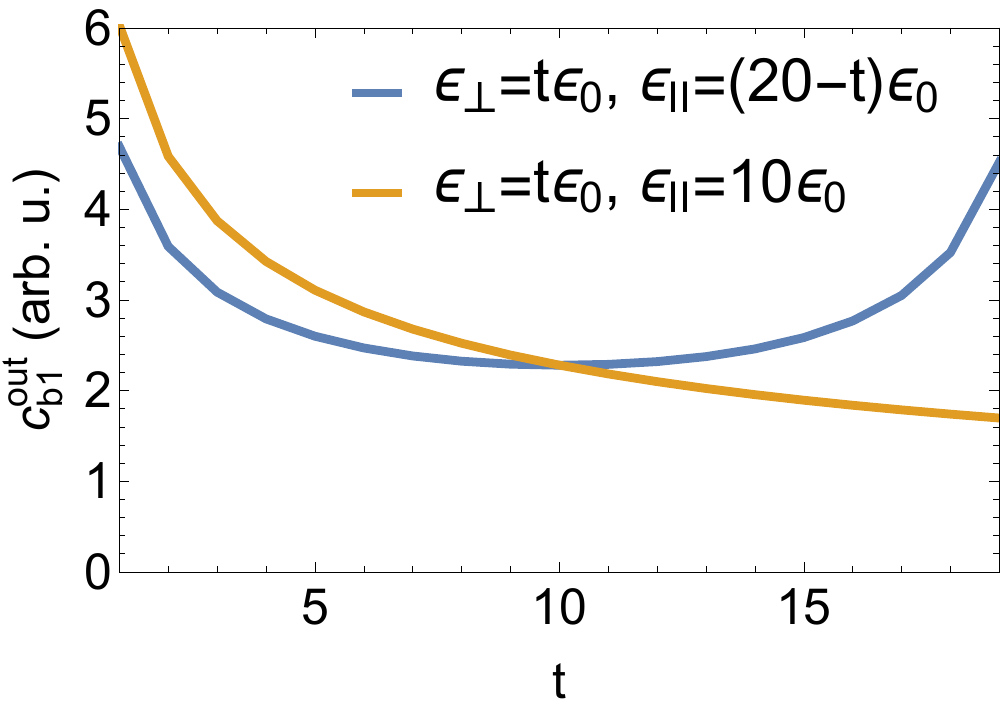}\end{minipage} 
\begin{minipage}{0.9\columnwidth}\hspace{-150pt} c)\\\includegraphics[width=0.6\columnwidth]{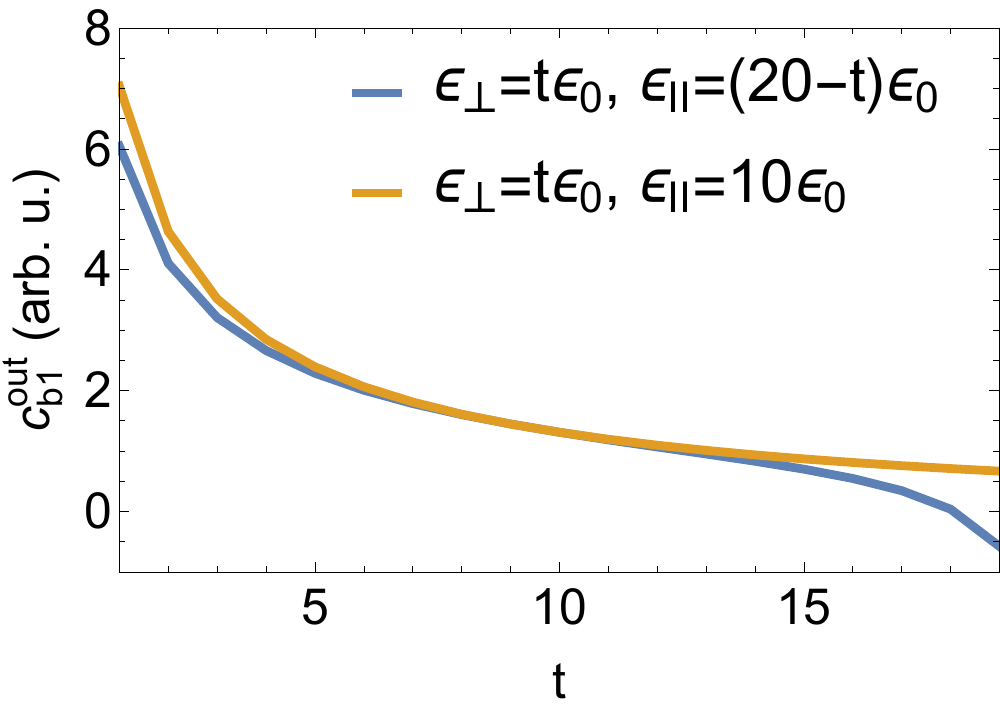}\end{minipage}
\caption{\label{fig3} (a) Strength of the monopolar field as a function of scaling $\alpha_\parallel$ (blue) and scaling $\alpha_\parallel$ by keeping $\alpha_\parallel+\alpha_\perp=const.$ (orange). In addition we show the evolution of monopolar strenght by a uniaxial scaling of the dielectric constant for a isotropic (b) magnetoelectric response $\alpha_\perp=\alpha_\parallel$  and for a strongly anisotropic (c) magnetoelectric response $\alpha_\perp=-3\alpha_\parallel$.}
\end{figure}

First, we investigate the dependence  of the magnetoelectric response on the anisotropy in $\alpha_\|$ and $\alpha_\perp$, with $\mat{\epsilon}$ and $\mat{\mu}$ set equal to isotropic values. In Fig.~\ref{fig3} (a) we show $c^{out}_{b1}$ as a function of $t$, which is the scaling between $\alpha_\|$ and $\alpha_\perp$, such that $\alpha_\perp=t~\alpha_\|$ for fixed $\alpha_\|$. We see that the monopolar field grows linearly with $\alpha_\|$ and vanishes for $\alpha_\|=-\alpha_\perp$. The orange line shows the change in monopole on keeping the sum of the components constant but varying the weight, thus $\alpha_\perp=t\alpha_0,\alpha_\|=(1-t)\alpha_0$. Interestingly, here the monopolar field strength remains independent of $t$, indicating that it is determined by the sum of both components rather than their relative magnitudes.

Next, we discuss the effect of the permittivity tensor on the field strength (the dependence on the permeability is analogous and we do not show it here), with the magnetoelectric response set to a isotropic value. In Fig.~\ref{fig3} b) (orange line) we plot the change in $c^{out}_{b1}$  when we linearly increase the perpendicular component $\epsilon_\perp=t\epsilon_0$  while keeping $\epsilon_\parallel$ constant. We see that the strength of the monopolar field decreases when $\epsilon$ increases. This is because a higher dielectric screening decreases the electric field inside the magnetoelectric which leads in turn to a reduced image monopole strength. With the blue line we show the result of setting $\epsilon_\perp=t\epsilon_0$ and $\epsilon_\|=(20-t)\epsilon_0$. This illustrates that the monopolar field is at its minimum for an isotropic tensor $\epsilon$, while a higher dielectric anisotropy increases the monopolar field regardless of which component of $\epsilon$ is increased.\\

Finally, in Fig.~\ref{fig3} c), we consider the situation in which we have anisotropy in both $\alpha$ and $\epsilon$, by setting $\alpha_\perp=-3\alpha_\parallel$ and varying $\epsilon$ in the same way as in  Fig.~\ref{fig3} b). In this case we find that increasing $\epsilon_\parallel$ leads to a reduced contribution of $\alpha_\parallel$ and vice versa.  Even the sign of the response can be changed if one element of $\epsilon$ is increased sufficiently, as is seen for values of $t>17$.

\clearpage

\section{LE-$\mu$-SR spectra}\label{app:musr}
Here we present representative raw data of our muon spectroscopy measurements presented in section \ref{sec:muonresults} for the example point with a stopping energy of 10 keV with an applied field of +10 mT.
\begin{figure}[h]
\includegraphics[scale=0.33]{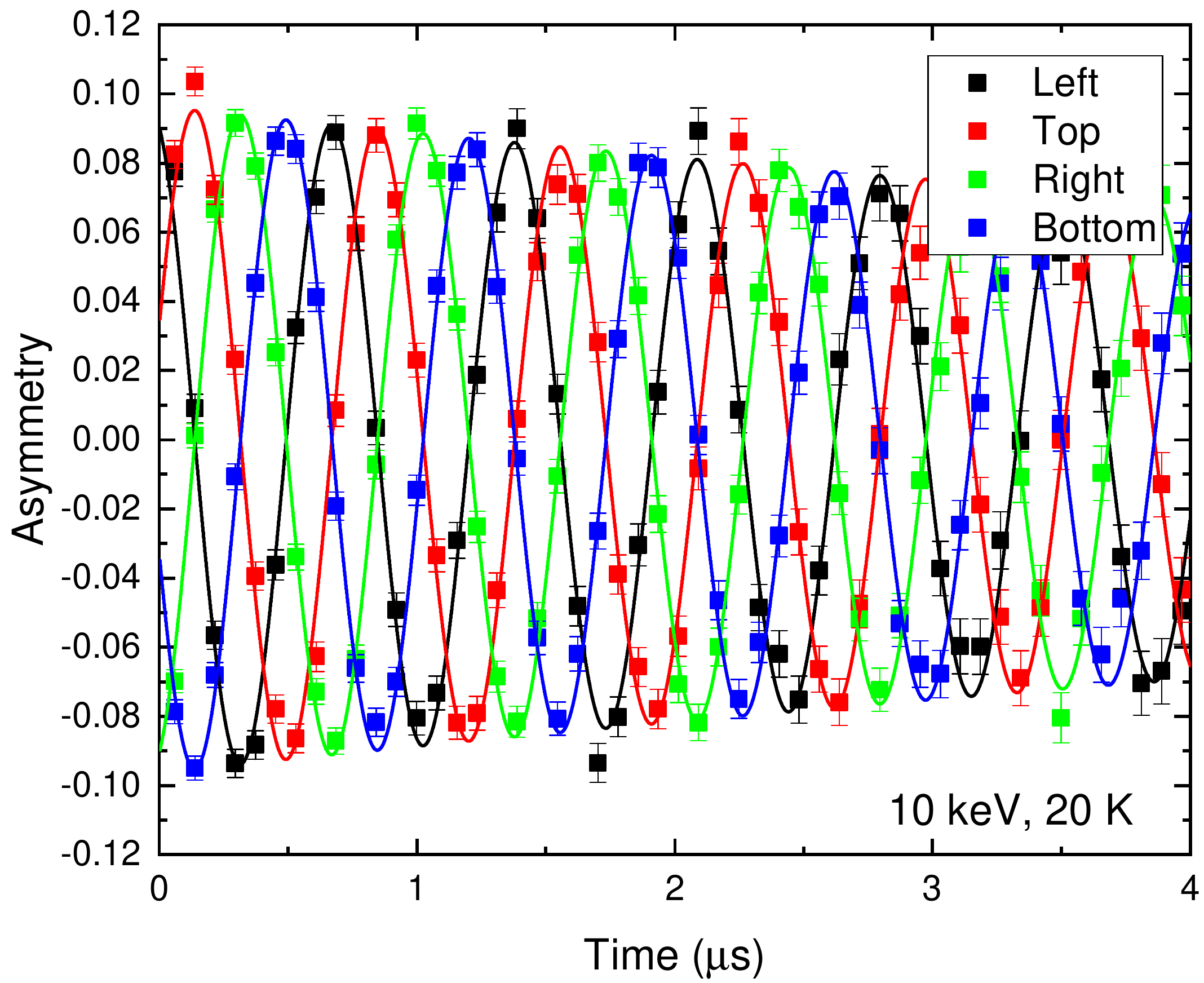}
\caption{Typical LE-$\mu$SR spectra (20 K, 10 keV, +10 mT) obtained for four positron detectors arranged around the sample. The solid lines are fits to the raw data with an exponential envelope function, i.e. assuming a Lorentzian field distribution.}\label{fig:spectra}
\end{figure}
\begin{figure}[h]
\includegraphics[scale=0.33]{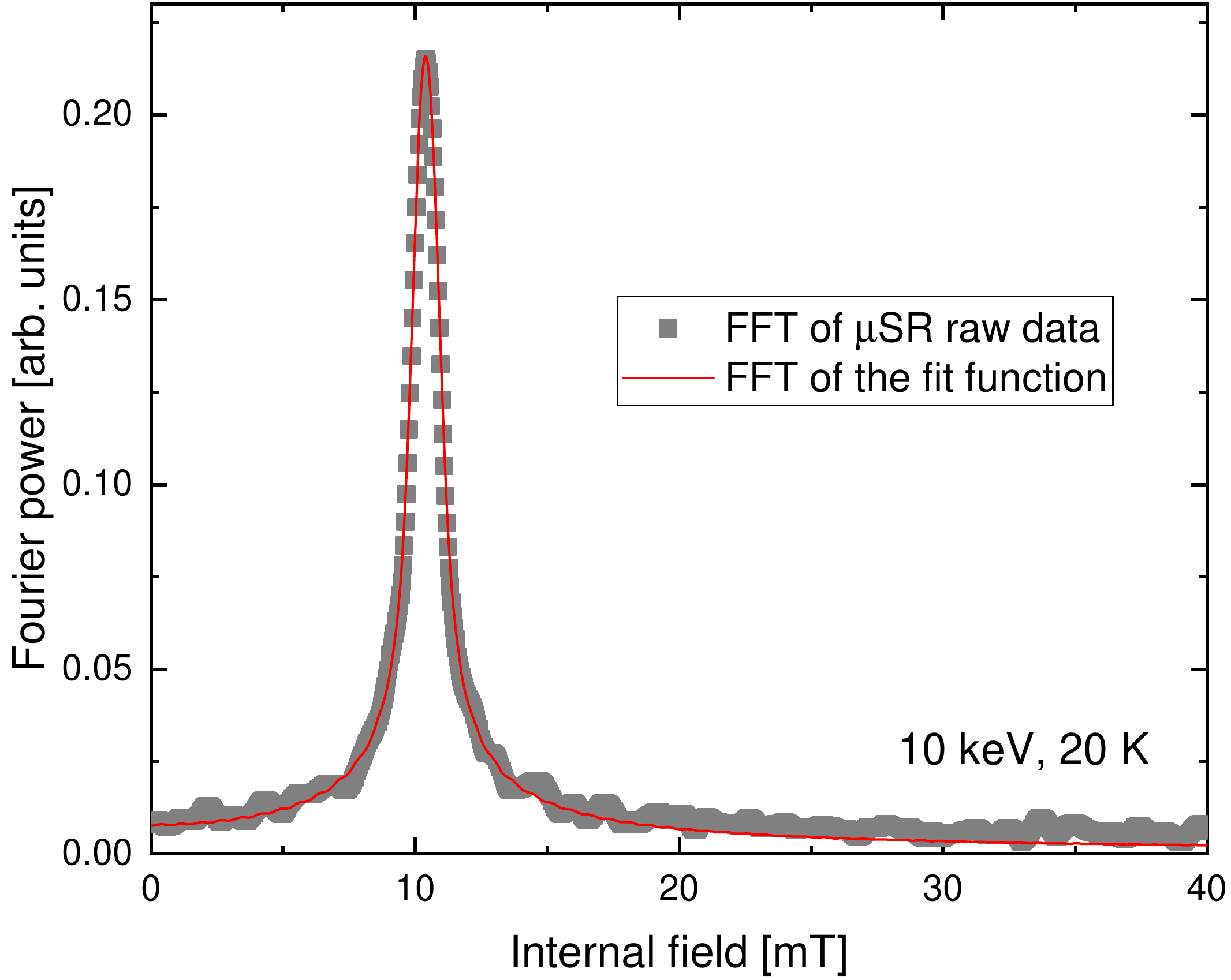}
\caption{Averaged FFT power of the $\mu$SR raw data (points) and of fits (solid line) shown in Fig. \ref{fig:spectra}. }
\end{figure}

\section{Magnetic flux through a SQUID loop}

Taking the usual form for the magnetic flux through a loop,
\begin{equation}
\Phi=\int\limits_{S}\mathbf{B}\cdot d\mathbf{S}
\end{equation}
and the magnetic field that we derived for a monopole at position $x=0$, $y=0$, $z=-d$,
\begin{equation}
\mathbf{B}=\dfrac{m}{(x^2+y^2+(z+d)^2)^{3/2}} \quad ,
\end{equation}
we integrate along the surface parametrized by
\begin{equation*}
\left\lbrace (x,y,z) \in S | x^2+y^2\leq R^2, z=z\right\rbrace
\end{equation*}
and obtain
\begin{align}
\Phi&=m\int\limits_S dS\dfrac{(z+d)}{(x^2+y^2+(z+d)^2)^{3/2}}\\
&=m(z+d)\int\limits_{0}^{2\pi}d\phi\int\limits_0^Rrdr \dfrac{1}{(r^2+(z+d)^2)^{3/2}} \quad .
\end{align}

Substituting $s=r^2+(z+d)^2$ and using $dr=\dfrac{ds}{2r}$ leads to
\begin{align}
\Phi&=2\pi\dfrac{m(z+d)}{2}\int\limits_{(z+d)^2}^{R^2+(z+d)^2}ds\left[\dfrac{1}{s^{3/2}}\right]\\&=2\pi m\left(1-\dfrac{z+d}{\sqrt{R^2+(z+d)^2}}\right) \quad .
\end{align}

Note that in the limit of a large loop radius, $R$, we find:
\begin{equation}
\Phi_{R\to\infty}=\dfrac{4\pi m}{2}
\end{equation} 
which is half the flux created by the point charge as expected from Gauss' Law.

\end{document}